\documentclass[prb,twocolumn,longbibliography,amsmath,superscriptaddress,amssymb]{revtex4-2}

\usepackage{graphicx}
\usepackage{dcolumn}
\usepackage{bm}
\usepackage{amssymb}
\usepackage{amsmath}
\usepackage{color}
\usepackage[dvipsnames]{xcolor}
\usepackage{placeins}
\usepackage{epstopdf}

\usepackage[normalem]{ulem}

\usepackage[linkcolor=blue,urlcolor=blue,citecolor=blue,colorlinks=true]{hyperref}

\newcommand{\spr}{\shortparallel}
\newcommand{\rtm}[1]{\mathrm{#1}}

\newcommand{\XU}{\mathbf{\hat x}}
\newcommand{\YU}{\mathbf{\hat y}}
\newcommand{\ZU}{\mathbf{\hat z}}

\newcommand{\ai}{{\it ab initio}}
\newcommand{\kp}{\hbox{$\mathbf{k}\cdot\mathbf{p}$}}

\begin{document}

\title{Plasmon manipulation by exchange magnetic field in two-dimensional spin-orbit coupled electronic systems: A higher-order relativistic $\mathbf{k}\cdot\mathbf{p}$ study}

\author{I. A. Nechaev}
\altaffiliation{Present address: Data Science and Digital Products Department, Viralgen, Paseo Mikeletegi 81, 20009 San Sebasti\'an, Spain}
\affiliation{Donostia International Physics Center (DIPC), Paseo Manuel de Lardizabal 4, 20018 Donostia/San Sebasti\'{a}n,  Basque Country, Spain}
\affiliation{Department of Electricity and Electronics, FCT-ZTF, UPV-EHU, 48080 Bilbao, Basque Country, Spain}

\author{V. M. Silkin}
\affiliation{Donostia International Physics Center (DIPC), Paseo Manuel de Lardizabal 4, 20018 Donostia/San Sebasti\'{a}n,  Basque Country, Spain}
\affiliation{Departamento de Pol\'{i}meros y Materiales Avanzados: F\'{i}sica, Qu\'{i}mica y Tecnolog\'{i}a, Universidad del Pa\'{i}s Vasco/Euskal Herriko Unibertsitatea, 20080 Donostia/San Sebasti\'{a}n, Basque Country, Spain}
\affiliation{IKERBASQUE, Basque Foundation for Science, 48013 Bilbao, Basque Country, Spain}

\author{E. E. Krasovskii}
\affiliation{Donostia International Physics Center (DIPC), Paseo Manuel de Lardizabal 4, 20018 Donostia/San Sebasti\'{a}n,  Basque Country, Spain}
\affiliation{Departamento de Pol\'{i}meros y Materiales Avanzados: F\'{i}sica, Qu\'{i}mica y Tecnolog\'{i}a, Universidad del Pa\'{i}s Vasco/Euskal Herriko Unibertsitatea, 20080 Donostia/San Sebasti\'{a}n, Basque Country, Spain}
\affiliation{IKERBASQUE, Basque Foundation for Science, 48013 Bilbao, Basque Country, Spain}

\date{\today}

\begin{abstract}
A higher-order relativistic \kp\ model is developed to describe plasmon excitations in two-dimensional (2D) electronic systems with spin–orbit coupling (SOC) and magnetic-exchange interactions. Derived entirely from \ai\ band structure, the model allows for a non-Rashba spin-momentum locking and enables a direct coupling of the exchange field to the real spin of electrons. Using the BiTeI trilayer (hexagonal $C_{3v}$ symmetry) and the Si-terminated surface state of TbRh$_2$Si$2$ (cubic $C_{4v}$ symmetry) as prototypes, we show that the exchange field induces strong, symmetry-dependent modifications of the band structure and plasmon dispersion. In BiTeI, it breaks the sixfold symmetry and leads to anisotropic, nonreciprocal plasmon modes, while in TbRh$_2$Si$_2$ it suppresses the characteristic triple spin winding and alters the plasmon damping. The results reveal that the interplay between SOC and exchange magnetism enables magnetic control of collective charge excitations in 2D spin-orbit systems beyond the Rashba paradigm.
\end{abstract}

\maketitle

\section{Introduction}
The spin-orbit interaction induced interplay of collective charge and spin excitations
is an important issue in two-dimensional (2D) spintronics. In the last decades, it has been
addressed for various model systems described by phenomenological Hamiltonians, in particular,
by the Rashba model of spin-orbit coupling for a trivial 2D electron gas
\cite{Chen_1999,Magarill2001,Ullrich2002,Wang_2005,Gumbs_2005,Kushwaha_2006,Pletyukhov_2006}
also including the Dresselhaus term~\cite{Ullrich2003,Badalyan2009,Agarwal2011,Maiti2015}
or specific low-energy effective Hamiltonians for helical liquids
\cite{Raghu2010,Efimkin2012,Roslyak2013,Stauber2013} also with the inclusion of a hexagonal
warping of the Fermi contour~\cite{LeBlanc_PRB_2014}. Plasmons in a spin-orbit coupled gas
in semiconducting quantum wells were considered based on the Luttinger Hamiltonian in
Ref.~\cite{Scholz2013} and on the Bernevig-Hughes-Zhang model in Ref.~\cite{Juergens2014}.
A common feature of these approaches is that the electron wave functions have a simplified
plane-wave form along the surface, which ignores the origin of the spin polarization being due
to the Coulomb singularities at the nuclei and limits the possible spin-momentum locking
geometries. Calculation beyond the low-energy continuum approximation were reported for
antimonene ~\cite{Slotman2018} and monolayer MoSi$_2$N$_4$~\cite{Wang2021} performed with
a tight-binding model.

Despite extensive studies of spin-orbit coupled 2D electron systems, the role of
the surface magnetization has not been sufficiently addressed, and the influence of magnetic
exchange fields on the collective charge excitations remains only partially understood.
To achieve a conclusive understanding, one must go beyond the low-order \kp\
Hamiltonians, and include higher-order relativistic corrections \cite{Nechaev_PRB_2018,Nechaev_PRB_2020}
responsible for the complex spin textures and warping in real materials. Avoiding heuristic simplifications is instrumental in a quantitative description of plasmonic phenomena in realistic spin-orbit systems, particularly in the presence of the exchange interaction. A comprehensive understanding of how the exchange field modifies dielectric screening and plasmon dispersion requires a theory that accurately captures both the spin-momentum locking and the anisotropy of the band structure. The challenge is to consistently derive the model from \ai\ electronic structure and thus ensure material-specific predictive power.

In this work, we develop a higher-order relativistic \kp\ model based on \ai\ wave
functions, which allows us to treat on the same footing the spin-orbit and exchange interactions in 2D systems of arbitrary symmetry. Using two representative cases---the conduction-band state of a BiTeI trilayer (hexagonal $C_{3v}$ symmetry) and the Rh-derived surface state at the Si-terminated surface of TbRh$_2$Si$2$ (cubic $C_{4v}$ symmetry)---we analyze how the exchange field of different orientation affects the electronic bands, spin textures, and collective charge excitations.

We show that the exchange field profoundly alters both the plasmon dispersion and its damping through symmetry-dependent modifications of the band structure and spin distribution. In BiTeI, an out-of-plane field reduces the sixfold rotational symmetry to threefold, while an in-plane field induces strongly anisotropic and nonreciprocal plasmon modes. In TbRh$_2$Si$_2$, the in-plane magnetization suppresses the characteristic triple winding of the in-plane spin and changes the selection rules for electronic transitions, leading to a substantial reorganization of the plasmon spectrum. These results demonstrate that the interplay between spin-orbit coupling and exchange magnetism provides an effective mechanism for tuning plasmon propagation, anisotropy, and nonreciprocity in spin-orbit coupled 2D systems.

\section{\label{two_band_model}Two-band $\mathbf{k}\cdot\mathbf{p}$ model}

In this section, we present a minimal $\mathbf{k}\cdot\mathbf{p}$ model for two-dimensional (2D) electronic systems with spin-orbit and magnetic-exchange interactions. The model is derived from the \ai\ band structure and does not resort to adjustable parameters. It is based on an effective Hamiltonian of high order in \textbf{k} and adequately describes a rich variety of spin structures arising in real systems. We employ the \kp\ model to show how the dielectric function of the 2D systems depends on band-structure variations brought about by spin-orbit and magnetic-exchange interactions.

\subsection{\label{EF_rep}Effective-field representation}

Consider a 2D electronic system described by the two-band Hamiltonian of the form
\begin{equation}\label{TwoBandHam}
  H_{\rtm{\mathbf{kp}}} = \mathcal{E}_{\mathbf{k}}\mathbb{I}_{2\times2} + {\bm{\mathcal{B}}}_{\mathbf{k}} \cdot{\bm\sigma}.
\end{equation}
Here, $\mathbb{I}_{2\times2}$ is the identity matrix and $\bm\sigma=(\sigma_x, \sigma_y, \sigma_z)$ is the vector of the Pauli matrices acting in a \textit{pseudospin} space. In Eq.~(\ref{TwoBandHam}), $\mathcal{E}_{\mathbf{k}}$ are the $\mathbf{k}$-dependent energies of the doubly degenerate states and $\bm{\mathcal{B}}_{\mathbf{k}}$ is an effective (spin-orbit) magnetic field that lifts the spin degeneracy. This field is introduced to simplify the representation and interpretation of the effect of spin-orbit interaction (SOI). Note that the vector field  $\bm{\mathcal{B}}_{\mathbf{k}}$ cannot be reduced to an expectation value of a vector operator such as, e.g., orbital angular momentum $\widehat{\mathbf{L}}$, total angular momentum $\widehat{\mathbf{J}}$ or spin $\widehat{\mathbf{S}}$.

Let us introduce the polar $\Theta_{\mathbf{k}}$ and azimuthal $\Phi_{\mathbf{k}}$ angles of the unit vector of the field:
\begin{equation}\label{unit_b}
{\hat{\mathbf{b}}}_{\mathbf{k}} = \frac{{\bm{\mathcal{B}}}_{\mathbf{k}} }{ | {\bm{\mathcal{B}}}_{\mathbf{k}} | }
= (\sin\Theta_{\mathbf{k}}\cos\Phi_{\mathbf{k}} ,\,  \sin\Theta_{\mathbf{k}} \sin\Phi_{\mathbf{k}} , \, \cos\Theta_{\mathbf{k}} ).
\end{equation}
The angles $\Theta_{\mathbf{k}}$ and $\Phi_{\mathbf{k}}$ are also the polar and azimuthal angles of a \textit{pseudospin}, since the effective field drives the \textit{pseudospin} to be collinear with ${\bm{\mathcal{B}}}_{\mathbf{k}}$ at a given ${\mathbf{k}}$. The explicit form of $\mathcal{E}_{\mathbf{k}}$ and $\bm{\mathcal{B}}_{\mathbf{k}}$ depends on the system and on the order of the $\mathbf{k}\cdot\mathbf{p}$ perturbation expansion, see Sec.~\ref{Field_details}.

The 2D vectors $\mathbf{C}^{\lambda}_{\mathbf{k}}=\left( C_{\uparrow \mathbf{k}}^{\lambda}, C_{\downarrow \mathbf{k}}^{\lambda} \right)^{\mathrm{T}}$ diagonalize the Hamiltonian $H_{\rtm{\mathbf{kp}}}$, $H_{\rtm{\mathbf{kp}}} \mathbf{C}^{\lambda}_{\mathbf{k}} = E^{\lambda}_{\mathbf{k}} \mathbf{C}^{\lambda}_{\mathbf{k}}$, with the energy eigenvalues
\begin{equation}\label{HamEig}
E^{\lambda}_{\mathbf{k}} = \mathcal{E}_{\mathbf{k}} + \lambda |\bm{\mathcal{B}}_{\mathbf{k}}|,
\end{equation}
where $\lambda=\pm$ labels the branches of the spin-orbit split band. The vectors $\mathbf{C}^{\lambda}_{\mathbf{k}}$ determine the state
\begin{equation}\label{model_state}
  |\widetilde{\Psi}^{\lambda}_{\mathbf{k}}\rangle = C_{\uparrow \mathbf{k}}^{\lambda} |\Psi_{\uparrow}\rangle + C_{\downarrow \mathbf{k}}^{\lambda} |\Psi_{\downarrow}\rangle
\end{equation}
in the reduced Hilbert space of the Hamiltonian of Eq.~(\ref{TwoBandHam}). Here, $|\Psi_{\mu}\rangle$ with $\mu=\uparrow$ or $\downarrow$ are the basis functions of the $\mathbf{k}\cdot\mathbf{p}$ model. In the \textit{ab initio} $\mathbf{k}\cdot\mathbf{p}$ theory of Refs.~\cite{Nechaev_PRBR_2016, Nechaev_PRB_2018, Nechaev_PRB_2019, Nechaev_PRB_2020, Nechaev_PRB_2021}, these are all-electron full-potential spinor wave functions found at a time-reversal invariant momentum (TRIM), and the subscript $\mu$ indicates the sign of the {\it on-site} expectation value of the $z$ component $\widehat{J}_z$ of the total angular momentum $\widehat{\mathbf{J}} = \widehat{\mathbf{L}} + \widehat{\mathbf{S}}$.  We express the vectors $\mathbf{C}^{\lambda}_{\mathbf{k}}$ in terms of the angles of the unit field vector ${\hat{\mathbf{b}}}_{\mathbf{k}}$ of Eq.~(\ref{unit_b}) as
\begin{equation}\label{C_vectors}
\mathbf{C}_{\mathbf{k}}^{+} = \left(
                                          \begin{array}{c}
                                            \cos\frac{\Theta_{\mathbf{k}}}{2} \\
                                            \sin\frac{\Theta_{\mathbf{k}}}{2}e^{i\Phi_{\mathbf{k}}} \\
                                          \end{array}
                                        \right),\,
\mathbf{C}_{\mathbf{k}}^{-} = \left(
                                          \begin{array}{c}
                                            -\sin\frac{\Theta_{\mathbf{k}}}{2}e^{-i\Phi_{\mathbf{k}}} \\
                                             \cos\frac{\Theta_{\mathbf{k}}}{2}\\
                                          \end{array}
                                        \right).
\end{equation}

The \textit{real spin} is obtained from the spin matrix $\rtm{\mathbf{S}}_{\rtm{\mathbf{kp}}} =(s^{\spr}\bm{\sigma}_{\spr}, s^{z}\sigma_z)$ with $\bm{\sigma}_{\spr}=(\sigma_x, \sigma_y)$. The elements of the spin matrix $[\rtm{\mathbf{S}}_{\rtm{\mathbf{kp}}}]^{\mu}_{\nu} = \langle\Psi_{\mu}|\bm{\sigma}|\Psi_{\nu}\rangle$ enter the expression for the spin expectation value in the eigenstate of the model of Eq.~(\ref{model_state})
%=====================================================================
\begin{eqnarray}\label{modelRealSpin}
\mathbf{S}_{\mathbf{k}}^{\lambda} = \frac{1}{2} \langle \widetilde{\Psi}^{\lambda}_{\mathbf{k}}|\bm{\sigma}|\widetilde{\Psi}^{\lambda}_{\mathbf{k}}\rangle
&=& \frac{1}{2}\sum\limits_{\mu\nu} C_{{\mathbf{k}}\mu}^{\lambda\ast}C_{{\mathbf{k}}\nu}^{\lambda} \left[\rtm{\mathbf{S}}_{\rtm{\mathbf{kp}}}\right]^{\mu}_{\nu} \nonumber \\
&=&\lambda \frac{1}{2}(s^{\spr} \hat{\mathbf{b}}^{\spr}_{\mathbf{k}} + s^{z} \hat{\mathbf{b}}^{z}_{\mathbf{k}}),
\end{eqnarray}
%=====================================================================
where $\hat{\mathbf{b}}^{\spr}_{\mathbf{k}}$ and $\hat{\mathbf{b}}^{z}_{\mathbf{k}}$ are the in-plane and out-of-plane components of the unit field vector defined in Eq.~(\ref{unit_b}), so ${\hat{\mathbf{b}}}_{\mathbf{k}} =  \hat{\mathbf{b}}^{\spr}_{\mathbf{k}} + \hat{\mathbf{b}}^{z}_{\mathbf{k}}$.

As follows from Eq.~(\ref{modelRealSpin}), in general, the \textit{real spin} and the effective field are non-collinear. However,  the in-plane spin $\mathbf{S}^{\spr \lambda}_{\mathbf{k}} = \lambda s^{\spr} \hat{\mathbf{b}}^{\spr}_{\mathbf{k}} / 2$ is always collinear with $\hat{\mathbf{b}}^{\spr}_{\mathbf{k}}$, and, therefore, for $s^{\spr}\neq0$ the locking angle between the momentum ${\mathbf{k}}$ and the spin $\mathbf{S}^{\spr \lambda}_{\mathbf{k}}$ is uniquely defined by the direction of $\hat{\mathbf{b}}^{\spr}_{\mathbf{k}}$, i.e., the in-plane component of the effective field at a given ${\mathbf{k}}$. For example, the classical Rashba model~\cite{Rashba_JETPL_1984, LaShell_PRL_1996} yields orthogonal spin-momentum locking. Following Ref.~\cite{Nechaev_PRB_2020}, we will characterize the non-orthogonality between $\mathbf{S}^{\spr \lambda}_{\mathbf{k}}$ and  ${\mathbf{k}}$ by the deviation angle $\delta^{\lambda}_{\mathbf{k}}$:
\begin{equation}\label{delta_nonorth}
\sin\delta^{\lambda}_{\mathbf{k}} = \frac{\mathbf{S}^{\spr \lambda}_{\mathbf{k}}\cdot \mathbf{k}} {|\mathbf{S}^{\spr \lambda}_{\mathbf{k}}|k}  = \hat{\mathbf{b}}^{\spr}_{\mathbf{k}}\cdot \hat{\mathbf{k}},
\end{equation}
where $\hat{\mathbf{k}}=\mathbf{k}/k$.

\subsection{\label{RPA_epsilon}Dielectric function}

Within the random phase approximation (RPA), the dielectric function
\begin{equation}\label{DF_RPA}
\epsilon({\mathbf q},\omega) = 1 - \frac{2\pi e^2}{|{\mathbf q}|} \chi_0({\mathbf q},\omega)
\end{equation}
of a spin-orbit coupled 2D system described by the Hamiltonian of Eq.~(\ref{TwoBandHam}) is completely defined by
the non-interaction charge-density response function
\begin{equation}\label{Chi0}
  \chi_0({\mathbf q},\omega) = \sum_{\lambda \lambda^{\prime}} \int \frac{d{\mathbf{k}}}{(2\pi)^2} \frac{(f_{\mathbf{k}}^{\lambda}-f_{\mathbf{k}+\mathbf{q}}^{\lambda^{\prime}})\mathcal{F}^{\lambda\lambda^{\prime}}_{\mathbf{k}, \mathbf{k}+\mathbf{q}}} {\omega + E^{\lambda}_{\mathbf{k}}-E^{\lambda^{\prime}}_{\mathbf{k}+\mathbf{q}}+ i\eta}
\end{equation}
with the Fermi factor $f_{\mathbf{k}}^{\lambda}$ and the factor
\begin{equation}\label{Overlapping}
\mathcal{F}^{\lambda\lambda^{\prime}}_{\mathbf{k}, \mathbf{k}+\mathbf{q}} = |\langle\widetilde{\Psi}^{\lambda}_{\mathbf{k}}|\widetilde{\Psi}^{\lambda^{\prime}}_{\mathbf{k}+ \mathbf{q}}\rangle |^2 =    \frac{1}{2}\left[1 + \lambda \lambda^{\prime} {\hat{\mathbf{b}}}_{\mathbf{k}} \cdot {\hat{\mathbf{b}}}_{\mathbf{k} + \mathbf{q}} \right]
\end{equation}
that accounts for the transition probability between the initial and final states $|\widetilde{\Psi}^{\lambda}_{\mathbf{k}}\rangle$ and $|\widetilde{\Psi}^{\lambda^{\prime}}_{\mathbf{k}+ \mathbf{q}}\rangle$. Note that the $\widetilde{\Psi}^{\lambda}$-states inherit their spatial structure from the all-electron basis spinors $|\Psi_{\mu}\rangle$, and the factor $\mathcal{F}^{\lambda\lambda^{\prime}} _{\mathbf{k}, \mathbf{k} +\mathbf{q}}$ turns out to depend only on the relative angle between the effective field (or equally \textit{pseudospin}) at the momenta ${\mathbf{k}}$ and ${\mathbf{k}+ \mathbf{q}}$, see also Ref.~\cite{Nechaev_PRB_2010}. This is an important point as it demonstrates that the transition amplitude depends on  the \textit{pseudospin} (the momentum-dependent effective field) rather than \textit{real spin}.

To analyze the effect of the external magnetic exchange field on the screening properties of the 2D systems, we construct the magnetic Hamiltonian $H^{\mathrm{M}}_{\rtm{\mathbf{kp}}} = H_{\rtm{\mathbf{kp}}} + H^{\rtm{EX}}$ as the sum of the Hamiltonian $H_{\rtm{\mathbf{kp}}}$ of Eq.~(\ref{TwoBandHam}) and the exchange term $H^{\rtm{EX}} = -\bm{\mathcal{J}} \cdot \rtm{\mathbf{S}}_{\rtm{\mathbf{kp}}}$, where $\bm{\mathcal{J}} = J_{\rtm{ex}} \mathbf{M}$ is a tunable parameter that allows for the magnetic exchange interaction of strength $J_{\rtm{ex}}$ with a magnetization $\mathbf{M}$. (Naturally, the Zeeman-like exchange term couples to the \textit{real spin}.) In other words, we model the magnetic phase by making the substitution $\bm{\mathcal{B}}_{\mathbf{k}}\rightarrow \widetilde{\bm{\mathcal{B}}}_{\mathbf{k}} = \bm{\mathcal{B}}_{\mathbf{k}} - s^{\spr}\bm{ \mathcal{J}}^{\spr} - s^{z} \bm{\mathcal{J}}^{z}$ in Eqs.~(\ref{TwoBandHam})--(\ref{Overlapping}).

\subsection{\label{Field_details}Symmetry and field constituents}

In this study we limit ourselves to hexagonal and cubic crystals with $C_{3v}$ and $C_{4v}$ symmetry, respectively. The $\mathbf{k}\cdot\mathbf{p}$ perturbation expansion is performed up to seventh order for the $C_{3v}$ and up to fifth order for the $C_{4v}$ case. The explicit momentum dependence of the two terms of the Hamiltonian of Eq.~(\ref{TwoBandHam})  is~\cite{Nechaev_PRB_2020, Usachov_PRL_2020}
\begin{eqnarray}
C_{3v}&&\left[\begin{array}{l}
         \mathcal{E}_{\mathbf{k}} = \mathcal{M}k^2+2\mathcal{N}k^6\cos6\varphi_{{\mathbf k}}, \\
         {\bm{\mathcal{B}}}_{\mathbf{k}} = \widetilde{\alpha}{\bm{\mathcal{B}}}_{\rtm{R}}^{(1)} + 2\mathcal{W}{\bm{\mathcal{B}}}_{\rtm{Z}}^{(3)} + \widetilde{\gamma}{\bm{\mathcal{B}}}^{(5)}_{+} +\xi{\bm{\mathcal{B}}}^{(7)};
       \end{array}\right.  \label{c3v_EandB} \\
       &\,& \nonumber \\
C_{4v}&&\left[\begin{array}{l}
         \mathcal{E}_{\mathbf{k}} = \mathcal{M}k^2+2\mathcal{N}k^4\cos4\varphi_{{\mathbf k}}, \\
        {\bm{\mathcal{B}}}_{\mathbf{k}} = \widetilde{\alpha}{\bm{\mathcal{B}}}_{\rtm{R}}^{(1)} + \mathcal{W}{\bm{\mathcal{B}}}_{\spr}^{(3)} + \widetilde{\gamma}{\bm{\mathcal{B}}}^{(5)}_{-},
       \end{array}\right. \label{c4v_EandB}
\end{eqnarray}
where all the coefficients except $\mathcal{N}$ and $\xi$ are a function of $k$:
\begin{eqnarray*}
\mathcal{M}&=&\sum_{m=0}^{n}\mathcal{M}^{(2m)}k^{2m}, \\
\mathcal{W}&=&\sum_{m=0}^{n}\mathcal{W}^{(2m)}k^{2m}, \\
\tilde{\alpha}&=&\sum_{m=0}^{n+1}\alpha^{(2m+1)}k^{2m}, \\ \tilde{\gamma}&=&\sum_{m=0}^{n-1}\gamma^{(2m+5)}k^{2m}
\end{eqnarray*}
with $n=1$ for $C_{4v}$ and $n=2$ for $C_{3v}$. Thus, in the present $\mathbf{k}\cdot\mathbf{p}$ theory a spin-orbit coupled 2D electronic system is characterized by a rather large set of the material-specific parameters. In the case of the $C_{4v}$ symmetry, these are $s^{\spr}$, $s^{z}$, $\mathcal{N}$, $\mathcal{M}^{(0)}$, $\mathcal{M}^{(2)}$, $\mathcal{W}^{(0)}$, $\mathcal{W}^{(2)}$, $\alpha^{(1)}$, $\alpha^{(3)}$, $\alpha^{(5)}$, and $\gamma^{(5)}$. For a system with the $C_{3v}$ symmetry, one must additionally specify the parameters $\mathcal{M}^{(4)}$, $\mathcal{W}^{(4)}$, $\alpha^{(7)}$,  $\gamma^{(7)}$, and $\xi$.

\begin{figure}[tbp]
\centering
\includegraphics[width=\columnwidth]{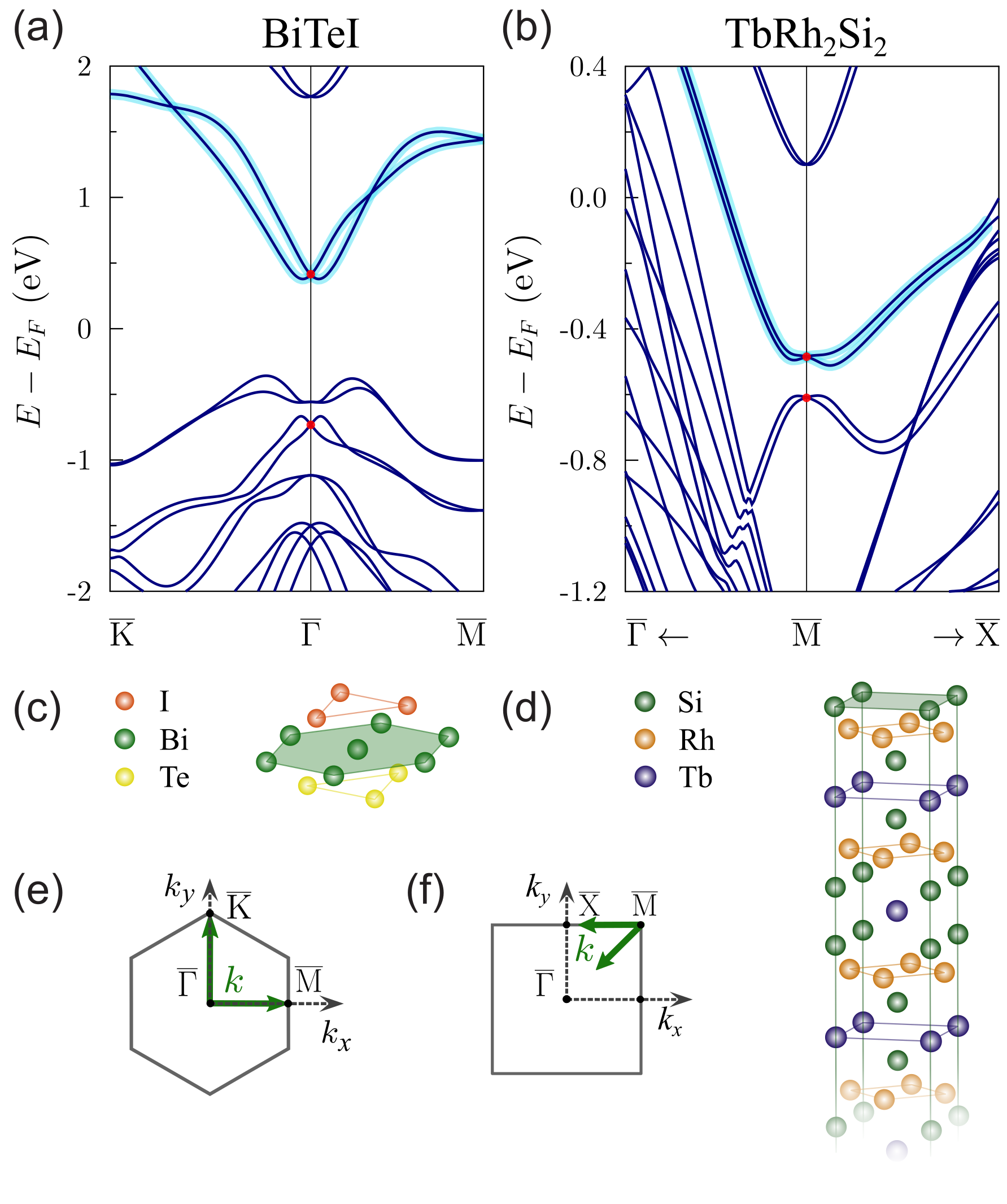}
\caption{Band structure of the BiTeI trilayer (a) and the Si-terminated centrosymmetric 31-layer slab (b) simulating the Si-terminated (001) surface of the paramagnetic TbRh$_2$Si$_2$.  The high-symmetry \textbf{k}-lines are shown in graphs (e) for BiTeI and (f) for TbRh$_2$Si$_2$ by green arrows. Highlighted in cyan are the lowest conduction-band state of the trilayer, graph (a), and the Rh-related higher-energy surface state, graph (b). These states serve as prototypes for the model 2D systems (see text). The atomic structure of the trilayer (c) and of the upper half of the 31-layer slab (d) are shown along with the corresponding 2D Brillouin zones (BZ) (e) and (f).}
\label{fig1}
\end{figure}

As prototypes of 2D electronic systems we take the lowest conduction-band state of a single trilayer of the hexagonal BiTeI, Fig.~\ref{fig1}(a), and the partly occupied higher-energy transition-metal surface state at the Si-terminated surface of the paramagnetic cubic TbRh$_2$Si$_2$, Fig.~\ref{fig1}(b). (In the following, we will refer to the respective prototypes as BTI and TRS models and to a pair of spin-orbit split 2D states as $\widetilde{\Psi}_{\bm{\rtm{k}}}^{\rtm{BTI}}$ and $\widetilde{\Psi}_{\bm{\rtm{k}}}^{\rtm{TRS}}$, respectively.) The material-specific parameters are obtained from the all-electron wave functions as explained in Ref.~\cite{Nechaev_PRB_2020}. The underlying \textit{ab initio} band structure is calculated with the extended linear augmented plane-wave method~\cite{Krasovskii_PRB_1997} for a centrosymmetric slab comprising two well-separated inversely stacked BiTeI trilayers~\cite{Nechaev_SciRep_2017} and a Si-terminated centrosymmetric 31-layer slab of TbRh$_2$Si$_2$~\cite{Usachov_PRL_2020}.

%+++++++++++++++++++++++++++++++++++++++++++++++++++++++++++++++++++++++++++++++++
%+++++++++++++++++++++++++++++++++++++++++++++++++++++++++++++++++++++++++++++++++

For the BiTeI trilayer, within the \textit{ab initio} third-order $\mathbf{k}\cdot\mathbf{p}$ perturbation expansion around $\bar{\Gamma}$, we generate a four-band Hamiltonian in the basis of two states: the lowest conduction-band state and the second highest valence-band state [see red points in Fig.~\ref{fig1}(a)]. Then, for the lowest conduction state, the two-band model of Eq.~(\ref{TwoBandHam}) with $\mathcal{E}_{\mathbf{k}}$ and ${\bm{\mathcal{B}}}_{\mathbf{k}}$ of Eq.~(\ref{c3v_EandB}) is analytically derived from the four-band Hamiltonian by means of the L\"{o}wdin partitioning, keeping the terms to seventh order in \textbf{k}. For the TbRh$_2$Si$_2$ surface, the expansion is performed around the $\bar{M}$ point, and the basis states for a four-band Hamiltonian comprise two Rh-related surface states [red points in Fig.~\ref{fig1}(b)]. The two-band model of Eq.~(\ref{TwoBandHam}) derived for the higher-energy Rh-related surface state includes the terms to fifth order in \textbf{k} [the terms of Eq.~(\ref{c4v_EandB})]. All the parameters obtained within this procedure are listed in Table~\ref{tab:table1}.

The basis sets of the four-band Hamiltonians are chosen so as to reproduce the \textit{ab initio} electronic structure of the states of interest, highlighted in cyan in Fig.~\ref{fig1}. For BTI, the choice of the basis was thoroughly discussed in  Ref.~\cite{Nechaev_SciRep_2017}. In TRS, we are interested in Rh-related surface state, so the basis is restricted to the states originating from the subsurface Rh atoms, which dominate the surface band and reproduce both the band warping and the characteristic triple spin winding, see Ref.~\cite{Usachov_PRL_2020}.

\begin{table}
\caption{\label{tab:table1} Parameters of the two-band Hamiltonians~(\ref{TwoBandHam}) for the BTI ($C_{3v}$ symmetry) and TRS ($C_{4v}$ symmetry) models. All parameters are in Rydberg atomic units: $\hbar=2m_0=e^2/2=1$.}
\begin{ruledtabular}
\begin{tabular}{ldd}
                                                    & \multicolumn{1}{c}{BTI}     &  \multicolumn{1}{c}{TRS} \\
\hline
$\alpha^{(1)}$                         &    0.238                     &   0.039          \\
$\alpha^{(3)}$                         & -41.671                    &    -11.850      \\
$\alpha^{(5)}$                         &  -167.006                 &  -925.401     \\
$\alpha^{(7)}$                         &  -159828.323          &                        \\
$\gamma^{(5)}$                      & 1176.607                 &   1264.246    \\
$\gamma^{(7)}$                      &  12763.879              &                         \\
$\xi$                                           &  -3125.750               &                         \\
$\mathcal{N}$                          &  -2748.111               &     -1435.192 \\
$\mathcal{M}^{(0)}$               &   5.353                      &     -1.550         \\
$\mathcal{M}^{(2)}$               &   -333.613                &    3888.197    \\
$\mathcal{M}^{(4)}$               & 114257.039            &                          \\
$\mathcal{W}^{(0)}$               &  9.133                       &     -23.350       \\
$\mathcal{W}^{(2)}$               &  1438.451                &      2294.544   \\
$\mathcal{W}^{(4)}$               &   119792.141          &                           \\
$s^{\spr}$                                   &    -0.47                      &  -0.99                \\
$ s^{z}$                                        &     -0.06                     &   0.99                \\
\end{tabular}
\end{ruledtabular}
\end{table}

The partial effective fields in the Taylor expansions in Eqs.~(\ref{c3v_EandB}) and (\ref{c4v_EandB}) are
\begin{eqnarray*}
  {\bm{\mathcal{B}}}_{\rtm{R}}^{(1)} &=& k(\sin\varphi_{{\mathbf k}},-\cos\varphi_{{\mathbf k}},0), \\
  {\bm{\mathcal{B}}}_{\rtm{Z}}^{(3)} &=& k^3(0,0,\sin3\varphi_{{\mathbf k}}), \\
  {\bm{\mathcal{B}}}_{\spr}^{(3)} &=& k^3(\sin3\varphi_{{\mathbf k}},\cos3\varphi_{{\mathbf k}},0), \\
 {\bm{\mathcal{B}}}^{(5)}_{\pm} &=& k^5(\sin5\varphi_{{\mathbf k}},\pm\cos5\varphi_{{\mathbf k}},0), \\
  {\bm{\mathcal{B}}}^{(7)} &=& k^7(\sin7\varphi_{{\mathbf k}}, - \cos7\varphi_{{\mathbf k}},0),
\end{eqnarray*}
where $\varphi_{{\mathbf k}}$ is the azimuth of ${\mathbf k}$ in the momentum plane. In the $C_{3v}$ case, apart from the classical Rashba field ${\bm{\mathcal{B}}}_{\rtm{R}}^{(1)}$, which gives rise to an axially symmetric spin structure with an orthogonal locking between ${\mathbf k}$ and the in-plane spin~\cite{Rashba_FTT_1959, Rashba_JETPL_1984, LaShell_PRL_1996}, the present $\mathbf{k}\cdot\mathbf{p}$ model includes both the third-order $z$-directed field ${\bm{\mathcal{B}}}_{\rtm{Z}}^{(3)}$ responsible for the well-known three-fold symmetric pattern of the out-of-plane spin~\cite{Fu_PRL_2009, Basak_PRB_2011} and the higher-order six-fold symmetric fields ${\bm{\mathcal{B}}}^{(5)}_{+}$ and ${\bm{\mathcal{B}}}^{(7)}$ allowing for the deviation of the locking angle from 90$^\circ$ with a $\pi/3$ periodicity as a function of $\varphi_{{\mathbf k}}$~\cite{Basak_PRB_2011, Nechaev_PRB_2020}. In the $C_{4v}$ case, the classical Rashba field is accompanied by the third-order four-fold symmetric field ${\bm{\mathcal{B}}}_{\spr}^{(3)}$ giving rise to the triple winding of in-plane spin along the constant energy contours (CECs)~\cite{Nechaev_PRB_2018, Schulz_QM_2019, Usachov_PRL_2020} and by the four-fold symmetry field ${\bm{\mathcal{B}}}^{(5)}_{-}$ also contributing to the non-Rashba behavior of the locking angle~\cite{Nechaev_PRB_2020}. Note that the presence of the third- and higher-order fields in ${\bm{\mathcal{B}}}_{\mathbf{k}}$ causes a modification of the scalar-relativistic warping of CECs due to the $\mathcal{N}$-factor term in Eqs.~(\ref{c3v_EandB}) and (\ref{c4v_EandB}) of $\mathcal{E}_{\mathbf{k}}$~\cite{Fu_PRL_2009, Basak_PRB_2011, Nechaev_PRB_2020}.

%+++++++++++++++++++++++++++++++++++++++++++++++++++++++++++++++++++++++++++++++++
\begin{figure*}[tbp]
\centering
\includegraphics[width=\textwidth]{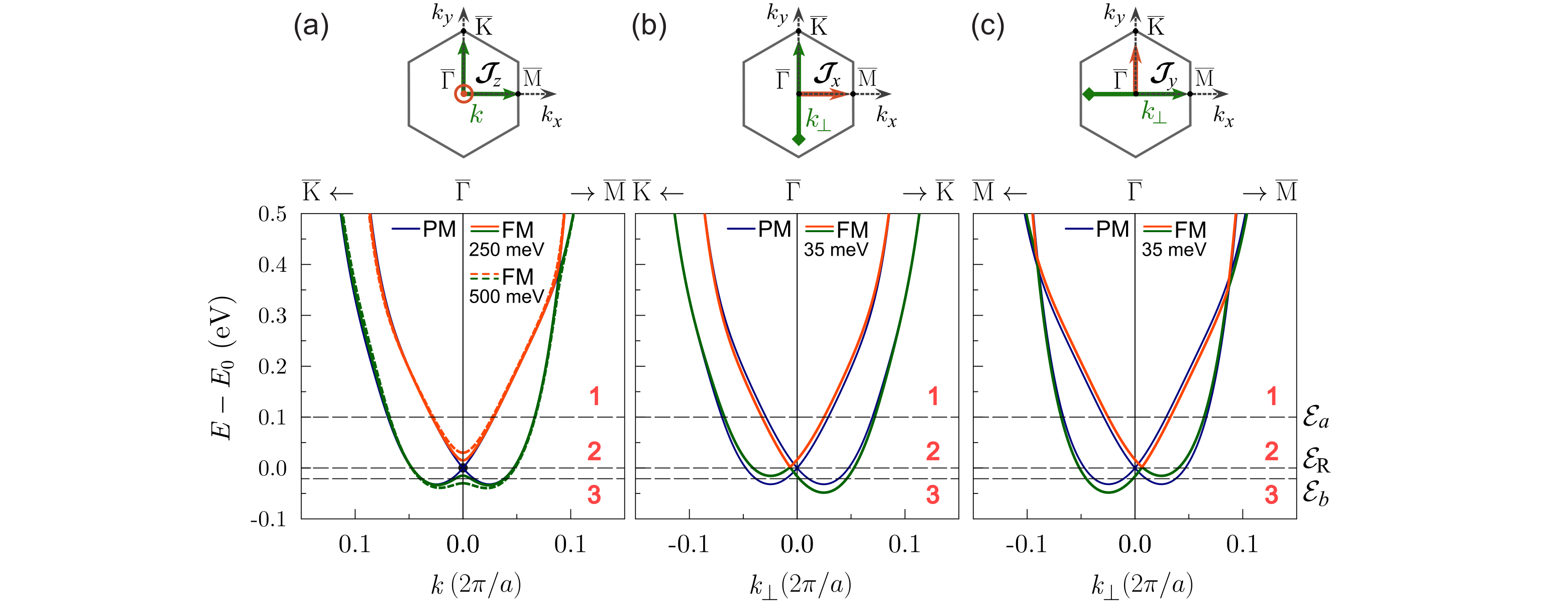}
\caption{Upper row: Green arrows show the directions in the 2D BZ along which the
eigenenergies in the lower row are presented. Red arrows show the direction of the exchange
field $\bm{\mathcal{J}}$. Lower row: Band structure of BTI by the two-band $\mathbf{k}\cdot\mathbf{p}$ Hamiltonian with the parameters listed in Table~\ref{tab:table1} for the PM and FM phases with the exchange field along $\ZU$~(a), $\XU$~(b), and $\YU$~(c). The respective values of the exchange parameter $\mathcal{J}$ are shown in the graphs. The green and orange lines correspond to the outer and inner branches of the split 2D states, respectively. The red numbers label the the positions  $\mathcal{E}_{a}$, $\mathcal{E}_{\rtm{R}}$, and $\mathcal{E}_{b}$ of the Fermi level $\mathcal{E}_{\mathrm{F}}$ for which the calculations of the dielectric function will be presented in the following.
}
\label{fig2}
\end{figure*}
%+++++++++++++++++++++++++++++++++++++++++++++++++++++++++++++++++++++++++++++++++

\section{\label{bands_and_CECs}Electronic bands and spin structures}

We now discuss the energy-momentum dispersion and spin-momentum locking in the model 2D systems. Within the minimal $\mathbf{k}\cdot\mathbf{p}$ model of Sec.~\ref{Field_details}, we show how these characteristics of the spin-orbit split states are affected by an external exchange field applied along different axes. To demonstrate the essential physics contained in the higher-order spin-orbit terms of the model Hamiltonian, we will analyze the nonorthogonality between in-plane spin and momentum.

\subsection{\label{bands_C3v}Hexagonal symmetry: BiTeI}

The band structure of BTI by the two-band Hamiltonian of Eq.~(\ref{TwoBandHam}) with $\mathcal{E}_{\mathbf{k}}$ and ${\bm{\mathcal{B}}}_{\mathbf{k}}$ of Eq.~(\ref{c3v_EandB}) and the parameters listed in Table~\ref{tab:table1} is shown in Fig.~\ref{fig2} for both the paramagnetic (PM) and ferromagnetic (FM) phase with the out-of-plane (along $\ZU$) and in-plane (along $\XU$ or $\YU$) orientations of the magnetization. Here, we consider three positions of the Fermi level $\mathcal{E}_{\rtm{F}}$ with respect to the Rashba point $\mathcal{E}_{\rtm{R}}$, see red numbers in Fig.~\ref{fig2}: (1)~above $\mathcal{E}_{\rtm{R}}$ ($\mathcal{E}_{\rtm{F}}=\mathcal{E}_{a}$), (2)~exactly at the Rashba point in the PM phase ($\mathcal{E}_{\rtm{F}}=\mathcal{E}_{\rtm{R}}$), and (3)~slightly below this point ($\mathcal{E}_{\rtm{F}}=\mathcal{E}_{b}$). We consider the out-of-plane magnetization with $\mathcal{J}=250$ and 500~meV and the in-plane magnetization in two directions with $\mathcal{J}=35$~meV.

In the absence of the exchange field (blue lines in Fig.~\ref{fig2}), at a nonzero momentum \textbf{k} the doubly degenerate state is spin-orbit split into two dispersion branches: the inner branch $\lambda=+$ and the outer branch $\lambda=-$, see Eq.~(\ref{HamEig}). These branches touch at $\bar{\Gamma}$, i.e., at the TRIM, where all effective fields are zero, and, in our particular case, also for the \textbf{k} points in the line $\bar{\Gamma}$-$\bar{M}$, where the effect of the isotropic Rashba field $\widetilde{\alpha}{\bm{\mathcal{B}}}_{\rtm{R}}^{(1)}$ is canceled by the combined action of the six-fold symmetric fields $\widetilde{\gamma}{\bm{\mathcal{B}}}^{(5)}_{+}$ and $\xi{\bm{\mathcal{B}}}^{(7)}$. The model system inherits from its \ai\ original the degeneracy of the branches away from $\bar{\Gamma}$, see Fig.~\ref{fig1}(a). At the same time, both near this degeneracy point and at higher energies the spin structure of BTI does not follow exactly the spin structure of the conduction-band state of the BiTeI trilayer. As seen in Fig.~\ref{fig2}(a), due to the non-zero spin parameter $s^z$, see Table~\ref{tab:table1}, the out-of-plane exchange field lifts the degeneracy, and the stronger the field the larger the gap between the branches.

%+++++++++++++++++++++++++++++++++++++++++++++++++++++++++++++++++++++++++++++++++
\begin{figure}[tbp]
\centering
\includegraphics[width=\columnwidth]{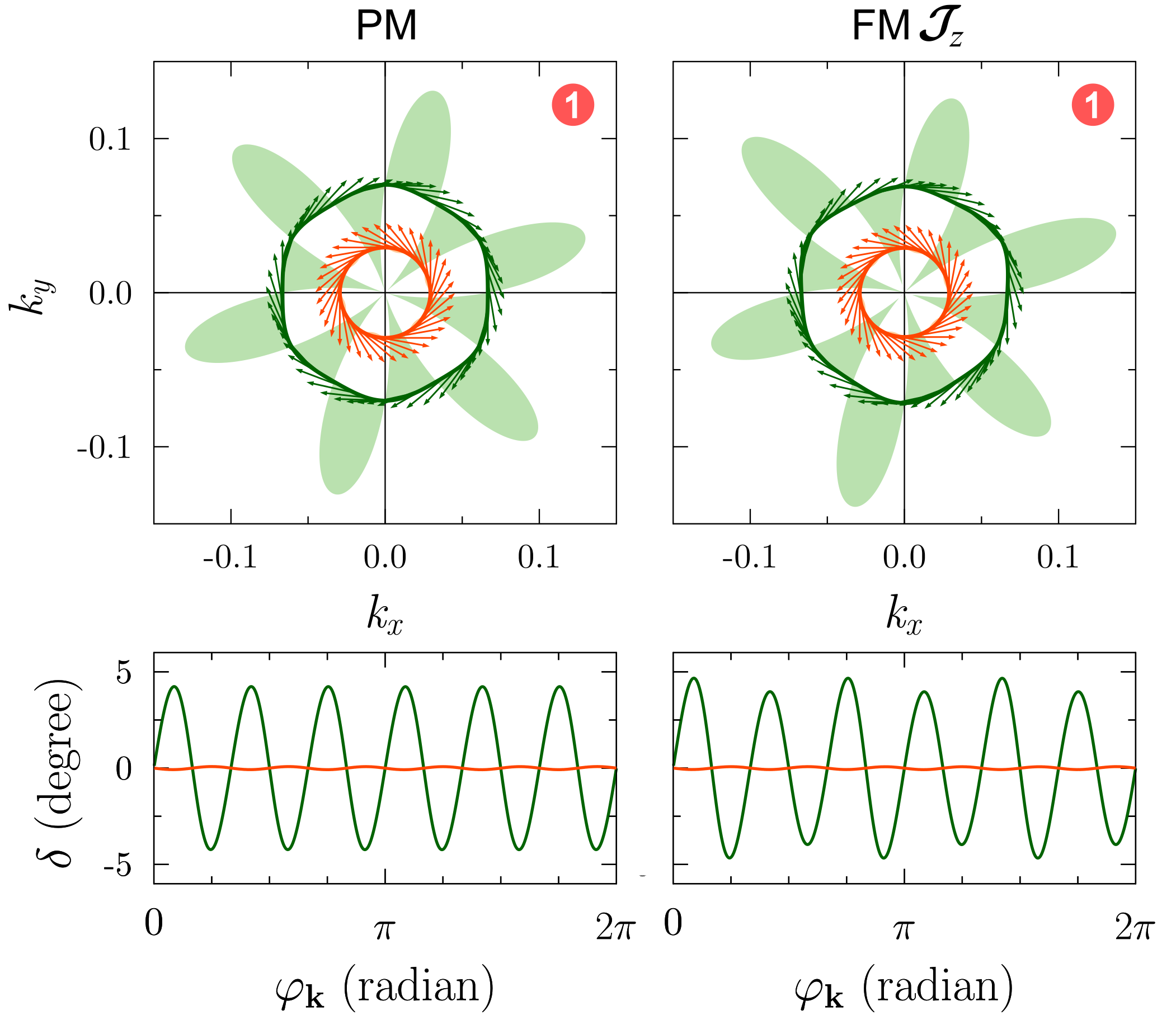}
\caption{Spin-resolved Fermi contours (upper panels) and the non-orthogonality $\delta^{\lambda}_{\mathbf{k}}$ defined in Eq.~(\ref{delta_nonorth}) as a function of $\varphi_{\mathbf{k}}$ (lower panels) for PM and FM BTI. In the FM phase, a $z$-directed exchange field with $\mathcal{J}_z = 250$~meV is implied. In the upper panels, the colored areas highlight the deviation of $\mathbf{S}^{\spr \lambda}_{\mathbf{k}}$ from the classical Rashba in-plane spin at a given $\mathbf{k}$-point in the contour: the border of the areas is given by $|\mathbf{k}|+R\sin\delta^{\lambda}_{\mathbf{k}}$ with $R$ being the scaling factor. The green areas and lines correspond to the outer branch of the split 2D state, while the orange ones to the inner branch. The contours are calculated at the Fermi energy $\mathcal{E}_{\mathrm{F}}$ indicated by 1 in Fig.~\ref{fig2}.}
\label{fig3}
\end{figure}
%+++++++++++++++++++++++++++++++++++++++++++++++++++++++++++++++++++++++++++++++++

As shown in the upper panels of Fig.~\ref{fig3}, for $\mathcal{E}_{\rtm{F}}=\mathcal{E}_{a}$ each branch of the split state gives rise to a closed Fermi contour. In the paramagnetic phase, both contours (inner and outer) have a six-fold symmetry, and the outer contour is much stronger warped than the inner one. The fields ${\bm{\mathcal{B}}}^{(5)}_{+}$ and ${\bm{\mathcal{B}}}^{(7)}$, which are responsible for the deviation from the orthogonal locking, also contribute to the hexagonal warping~\cite{Nechaev_PRB_2020}, so the stronger warping of the outer contour implies a larger nonorthogonality. This is clearly seen in the related lower panel of Fig.~\ref{fig3}: The calculated deviation angle $\delta^{\lambda}_{\mathbf{k}}$ defined in Eq.~(\ref{delta_nonorth}) behaves rather differently for the inner and outer Fermi contour. It oscillates as a function of $\varphi_{{\mathbf k}}$ with a $\pi/3$ periodicity and an amplitude around $0.1^{\circ}$ for the inner and $4.5^{\circ}$ for the outer contour.

Due to the presence of the field ${\bm{\mathcal{B}}}^{(3)}_{\rtm{Z}}$, the $z$-directed magnetic exchange field distorts the contours so they have a three-fold symmetry, Fig.~\ref{fig3}. Accordingly, the deviation angle $\delta^{\lambda}_{\mathbf{k}}$ now oscillates with a $2\pi/3$ periodicity. The effect of the out-of-plane field is seen to be more pronounced for the outer contour. It is noteworthy that within a third-order $\mathbf{k}\cdot\mathbf{p}$ model, which yields the orthogonal spin-momentum locking, the exchange field along $\ZU$ does not couple with the in-plane spin structure.

\begin{figure}[tbp]
\centering
\includegraphics[width=\columnwidth]{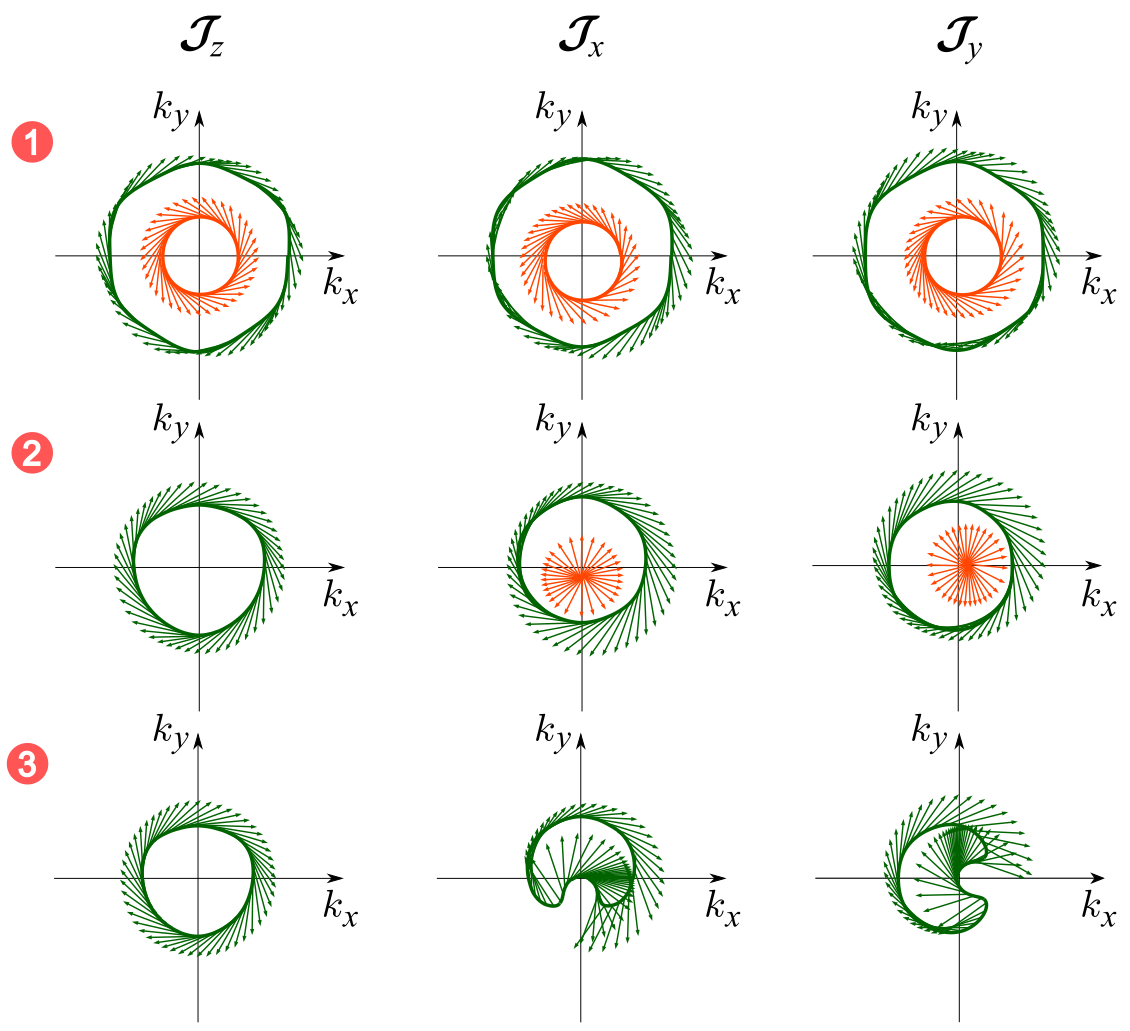}
\caption{Spin-resolved Fermi contours for the energies marked by red numbers in Fig.~\ref{fig2} for three FM phases of BTI. For the $z$-directed field, in the cases 1 and 2 the exchange-interaction parameter is $\mathcal{J}=250$~meV, while in the case 3 it is $\mathcal{J}=500$~meV. For the $x$- and $y$-directed fields, $\mathcal{J}=35$~meV in all cases.}
\label{fig4}
\end{figure}

The spin-resolved Fermi contours of $\widetilde{\Psi}_{\bm{\rtm{k}}}^{\rtm{BTI}}$ under the out-of-plane field for the three positions of the Fermi level are shown in the left column of Fig.~\ref{fig4}. The figure demonstrates that, in contrast to the $\mathcal{E}_{\rtm{F}}=\mathcal{E}_{a}$ case described above, for $\mathcal{E}_{\rtm{F}}=\mathcal{E}_{\rtm{R}}$ and $\mathcal{E}_{\rtm{F}}=\mathcal{E}_{b}$ there is only one Fermi contour because of the field-induced gap at $\bar{\Gamma}$, see Fig.~\ref{fig2}(a). The contour arises from the outer branch and has a three-fold symmetry. The deviation angle for this contour is not shown, since it is rather small, less than~$1^{\circ}$.

%+++++++++++++++++++++++++++++++++++++++++++++++++++++++++++++++++++++++++++++++++
%+++++++++++++++++++++++++++++++++++++++++++++++++++++++++++++++++++++++++++++++++

As in the spectrum of an ordinary Rashba-like 2D system, the in-plane exchange field shifts the Rashba point away from $\bar{\Gamma}$ in the direction perpendicular to the magnetization, Figs.~\ref{fig2}(b) and \ref{fig2}(c).  The exchange interaction also shifts the Fermi contours perpendicular to the field, with the inner and outer contours moving in opposite directions, see the case $\mathcal{E}_{\rtm{F}}=\mathcal{E}_{a}$  and $\mathcal{E}_{\rtm{F}}=\mathcal{E}_{\rtm{R}}$ in the middle and right columns of Fig.~\ref{fig4}. Moreover, depending on the direction of the in-plane field (along $\XU$ or $\YU$), it differently distorts the contours and modifies rather strongly the spin-momentum locking induced by the SOI. In contrast to the case of the magnetization along $\ZU$, here $\delta^{\lambda}_{\mathbf{k}}$ is not informative as a measure of the deviation from the in-plane spin structure of the Rashba model, since the in-plane exchange interaction by itself induces a nonorthogonal locking, which is treated differently in $\mathbf{k}\cdot\mathbf{p}$ models of different order in \textbf{k}.

%+++++++++++++++++++++++++++++++++++++++++++++++++++++++++++++++++++++++++++++++++
\begin{figure*}[tbp]
\centering
\includegraphics[width=\textwidth]{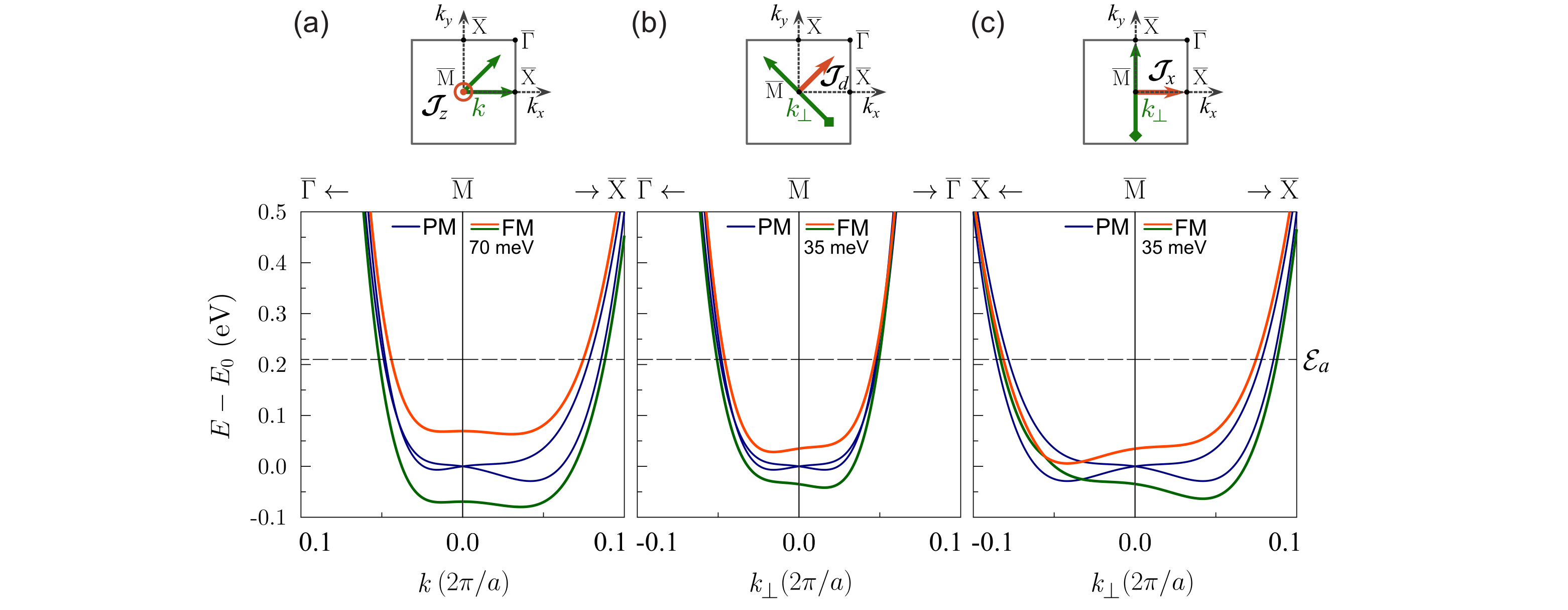}
\caption{Same as in Fig.~\ref{fig2} but for TRS. The exchange field is along $\ZU$~(a), along the diagonal $(\XU+\YU)/\sqrt{2}$~(b), and along $\XU$~(c).}
\label{fig5}
\end{figure*}
%+++++++++++++++++++++++++++++++++++++++++++++++++++++++++++++++++++++++++++++++++

As follows from Figs.~\ref{fig2}(b) and \ref{fig2}(c), the in-plane field also leads to the outer-branch minima having different energies on the opposite sides of $\bar{\Gamma}$. As a consequence, for $\mathcal{E}_{\rtm{F}} = \mathcal{E}_{b}$ the Fermi contour has an inwardly concave arc with the in-plane spin rotating in the opposite direction relative to the other parts of the contour, see the bottom of the middle and right columns of Fig.~\ref{fig4}.

The above brief analysis of the exchange-field effect on the spectrum and in-plane spin structure in BTI reveals that a relatively small Zeeman term $H^{\rtm{EX}}$ distorts tangibly the energy-momentum dispersion and spin-momentum locking of $\widetilde{\Psi}_{\bm{\rtm{k}}}^{\rtm{BTI}}$. In turn, this affects the phase space and energies of the intrabranch [$\lambda=\lambda^{\prime}$ in Eq.~(\ref{Chi0})] and interbranch [$\lambda\neq\lambda^{\prime}$] transitions and ultimately, as we show in Sec.~\ref{C3v_plasmons}, modify the plasmon dispersion and linewidth.

\subsection{\label{bands_C4v}Cubic symmetry: TbRh$_2$Si$_2$}
 Here we consider only one position of the Fermi level: the one above the Rashba point. Two magnetic states will be analyzed: an out-of-plane magnetization with $\mathcal{J}=70$~meV and an in-plane one with $\mathcal{J}=35$~meV. Unlike the true surface state of the Si-terminated TbRh$_2$Si$_2$, around the $\bar{M}$ point the spin-orbit split state $\widetilde{\Psi}_{\bm{\rtm{k}}}^{\rtm{TRS}}$ has no exotic spin polarization related to the vanishing spin expectation value in approaching the $\bar{M}$ point~\cite{Nechaev_PRB_2018, Usachov_PRL_2020}. Therefore, in TRS the effect of the in-plane exchange field differs from that in the real system, where the Rashba point remains almost unaffected by the field~\cite{GdRhSi, Nechaev_PRB_2018}.

\begin{figure}[tbp]
\centering
\includegraphics[width=\columnwidth]{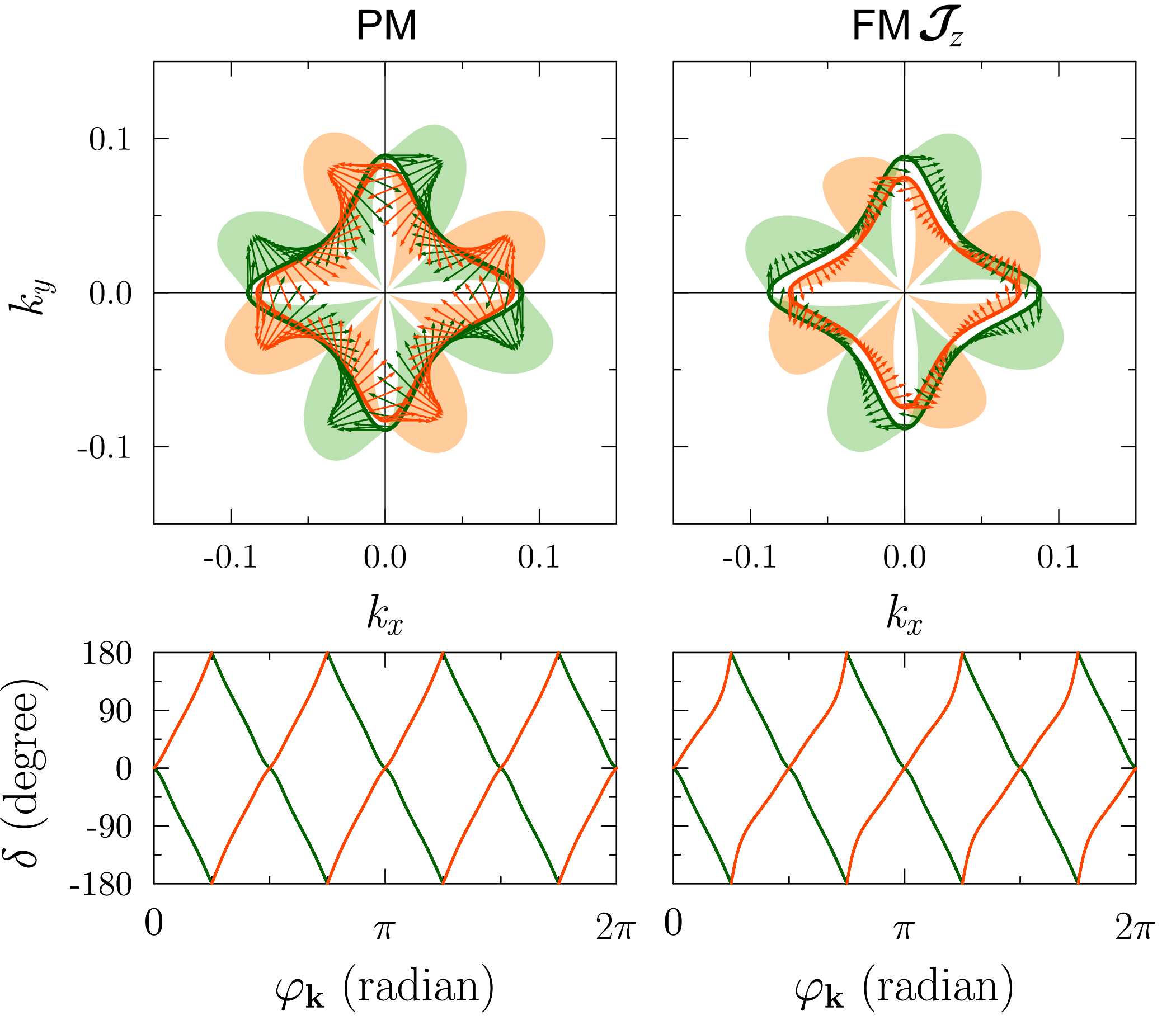}
\caption{Same as in Fig.~\ref{fig3} but for TRS. In the FM phase the exchange field is along $\ZU$ with $\mathcal{J}_z = 70$~meV.}
\label{fig6}
\end{figure}

The PM TRS model is characterized by a strongly anisotropic dispersion and spin-orbit splitting of the 2D state not only at a distance from the Rashba point but also in its vicinity [blue lines in Fig.~\ref{fig5}(a)]. At the Fermi level, this manifests itself in a strong four-fold warping of both inner and outer Fermi contour (as a quadrangular star shape) and the $\varphi_{{\mathbf k}}$-dependent $\mathbf k$-splitting between the contours in the momentum plane, see the upper-left panel of Fig.~\ref{fig6}. Note that the $\mathbf k$-splitting is small due to the weak SOI in TbRh$_2$Si$_2$.

The spin structure of TRS state $\widetilde{\Psi}_{\bm{\rtm{k}}}^{\rtm{TRS}}$ has the distinctive triple winding of the in-plane spin inherited from the true Si-terminated surface state~\cite{Usachov_PRL_2020}. This means that in moving around the Fermi contour the spin rotates three times faster than in the Rashba-type spin structures, with only a single winding. (The spin structure of $\widetilde{\Psi}_{\bm{\rtm{k}}}^{\rtm{BTI}}$ also manifests a single winding of the in-plane spin, albeit with a tangible deviation from the orthogonal locking, see the left column of Fig.~\ref{fig3}.) The faster rotation is clearly seen in the lower-left panel of Fig.~\ref{fig6}: the anticlockwise rotation of the \textbf{k} vector by $45^{\circ}$---and hence the Rashba-model spin---is accompanied by a change in the deviation angle by $180^{\circ}$ (see, e.g., $\delta^{\lambda}_ {\mathbf{k}}$ at $\varphi_ {{\mathbf k}}=\pi/4$), which corresponds to the clockwise rotation of the TRS-model spin by $135^{\circ}$.

Similar to $\widetilde{\Psi}_{\bm{\rtm{k}}}^{\rtm{BTI}}$, in TRS the out-of-plane exchange field creates a gap between the $\widetilde{\Psi}_{\bm{\rtm{k}}}^{\rtm{TRS}}$ branches at all \textbf{k}, Fig.~\ref{fig5}(a). (In spite of the smaller exchange-interaction parameter than in BTI, the exchange term $H^{\rtm{EX}}$ is larger due to the spin parameter $s^z$, which is almost unity here, and, therefore, the gap around $\bar{M}$ is larger too.) However, unlike BTI, under the out-of-plane field the Fermi contours of $\widetilde{\Psi}_{\bm{\rtm{k}}}^{\rtm{TRS}}$ have the same four-fold symmetry as in the paramagnetic TRS model, Fig.~\ref{fig6}, the main effect on the in-plane spin being a significant decrease in the spin magnitude, while the in-plane spin-momentum locking remains practically unaltered~\cite{Nechaev_PRB_2018, Usachov_PRL_2020}.

%+++++++++++++++++++++++++++++++++++++++++++++++++++++++++++++++++++++++++++++++++
%+++++++++++++++++++++++++++++++++++++++++++++++++++++++++++++++++++++++++++++++++

\begin{figure}[tbp]
\centering
\includegraphics[width=\columnwidth]{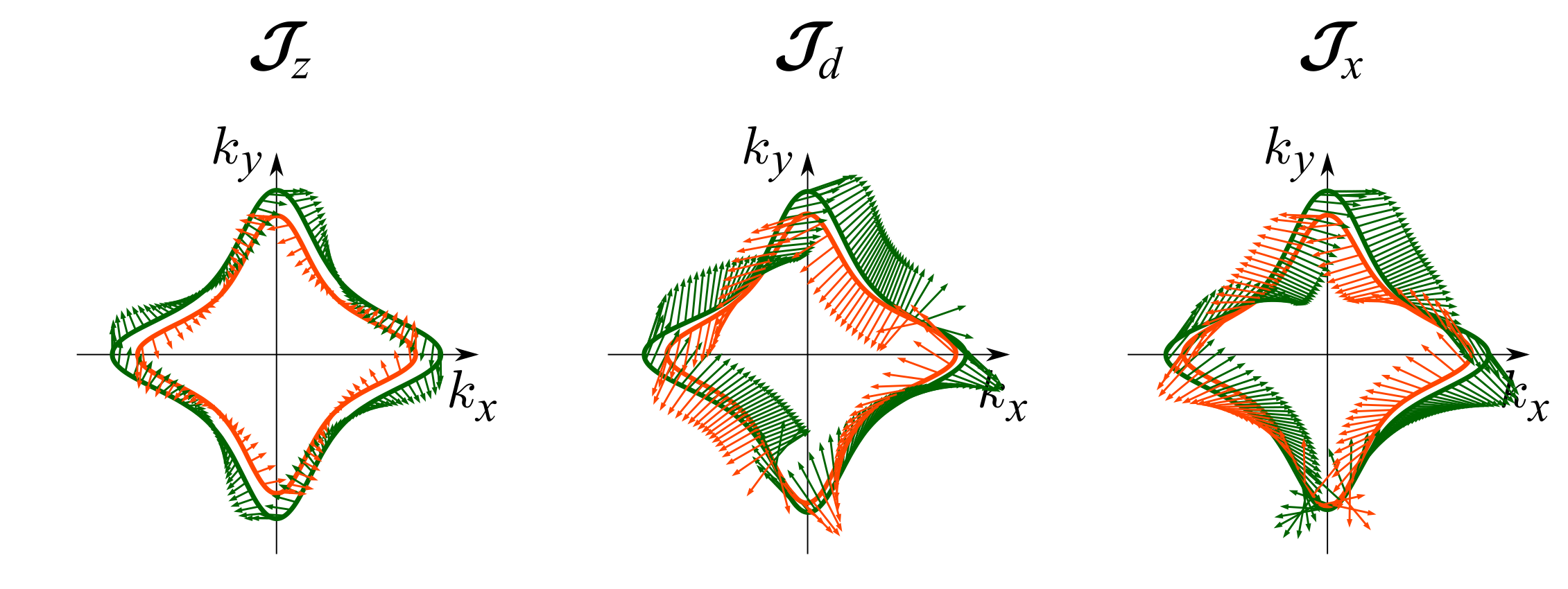}
\caption{Same as in Fig.~\ref{fig4} but for TRS. The Fermi level is shown in Fig.~\ref{fig5}. For the $z$-directed field, the exchange-interaction parameter $\mathcal{J}$ is 70~meV, while for the in-plane field it is 35~meV.}
\label{fig7}
\end{figure}

In the magnetic TRS model, the in-plane exchange field parameter is the same as in the BTI model with the in-plane magnetization. However, in TRS the exchange term $H^{\rtm{EX}}$ is nearly twice as large because $|s^{\spr}|$ is very close to unity. Along with the substantially weaker SOI, this leads to a stronger effect of the exchange field. Actually, as seen in Fig.~\ref{fig5}(b) and \ref{fig5}(c), the Rashba point moves far away from $\bar{M}$ perpendicular to the magnetization, and the splitting considerably changes. The pronounced distortion of the Fermi contours is accompanied by the rotation of the in-plane spin, so finally it strongly gravitates toward the field direction, Fig.~\ref{fig7}. As follows from the middle and right panel of Fig.~\ref{fig7}, the sign of the in-plane spin projection parallel to the field does not vary along the outer or inner contour, except some small arcs in the quadrangular-star rays where the inner and outer contours almost touch.

%+++++++++++++++++++++++++++++++++++++++++++++++++++++++++++++++++++++++++++++++++
%+++++++++++++++++++++++++++++++++++++++++++++++++++++++++++++++++++++++++++++++++

The unique spin-momentum locking of $\widetilde{\Psi}_{\bm{\rtm{k}}}^{\rtm{TRS}}$, which survives in the out-of-plane magnetic field but can be destroyed by an in-plane one, implies that the matrix element of Eq.~(\ref{Overlapping}) is radically different from that in a typical Rashba system. Along with the field-induced distortions of the highly anisotropic spectrum, this peculiarity manifests itself in the way the electron-hole continuum and plasmon dispersion is formed, as we analyze below in Sec.~\ref{C4v_plasmons}.

\section{\label{EELS}Electron energy-loss spectra}

In this section, for the BTI and TRS models we analyse the loss function $L(\mathbf{q}, \omega) = -\mathrm{Im}[1/\epsilon(\mathbf{q}, \omega)]$, where $\epsilon(\mathbf{q}, \omega)$ is the dielectric function of Eq.~(\ref{DF_RPA}) with the charge-density response function of Eq.~(\ref{Chi0}). We discuss the energy-loss spectra in terms of intrabranch (within a branch of the split state) and interbranch (between the branches) electron-hole excitations and their interplay with plasmon modes. We put a special focus on the effect of the exchange field on the dispersion and spectral width (damping) of the plasmon.

\subsection{\label{C3v_plasmons}Hexagonal symmetry: BiTeI}

%+++++++++++++++++++++++++++++++++++++++++++++++++++++++++++++++++++++++++++++++++
\begin{figure*}[tbp]
\centering
\includegraphics[width=\textwidth]{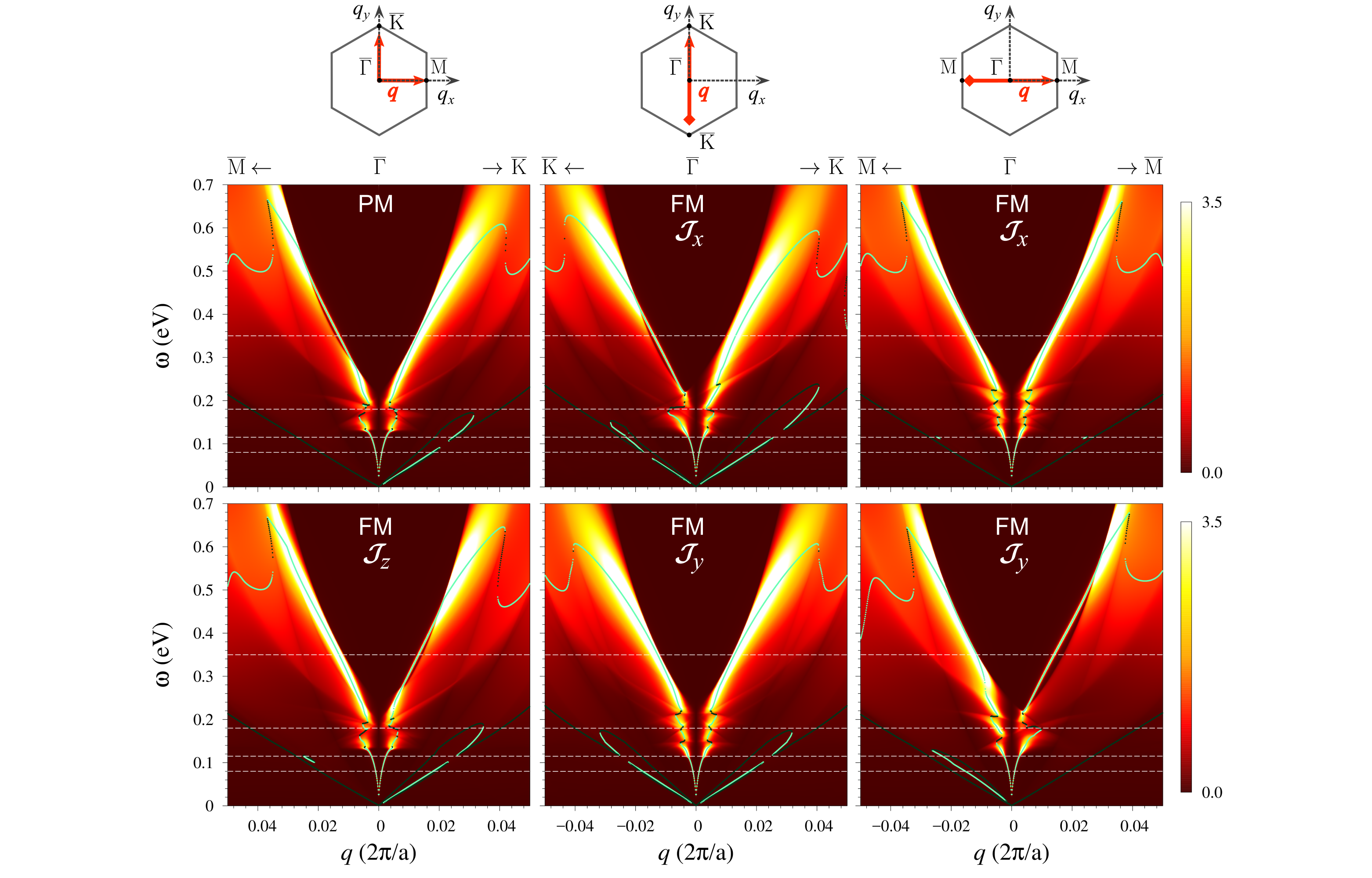}
\caption{Energy-loss spectra of BTI in the PM and FM phases with $\mathcal{E}_{\rtm{F}}=\mathcal{E}_{a}$ along the $\mathbf{q}$-lines indicated at the top of each column. White dashed lines show the energies at which the constant-$\omega$ cuts of the loss spectra are calculated, see Fig.~\ref{fig9}. Light and dark green points are zeros of the real part of the dielectric function: $\epsilon_1(\mathbf{q}, \omega)=0$. The shade of green encodes the sign of the energy derivative of $\epsilon_1$ at a given zero: light for positive and dark for negative derivative (see text).}
\label{fig8}
\end{figure*}
%+++++++++++++++++++++++++++++++++++++++++++++++++++++++++++++++++++++++++++++++++

\begin{figure*}[tbp]
\centering
\includegraphics[width=\textwidth]{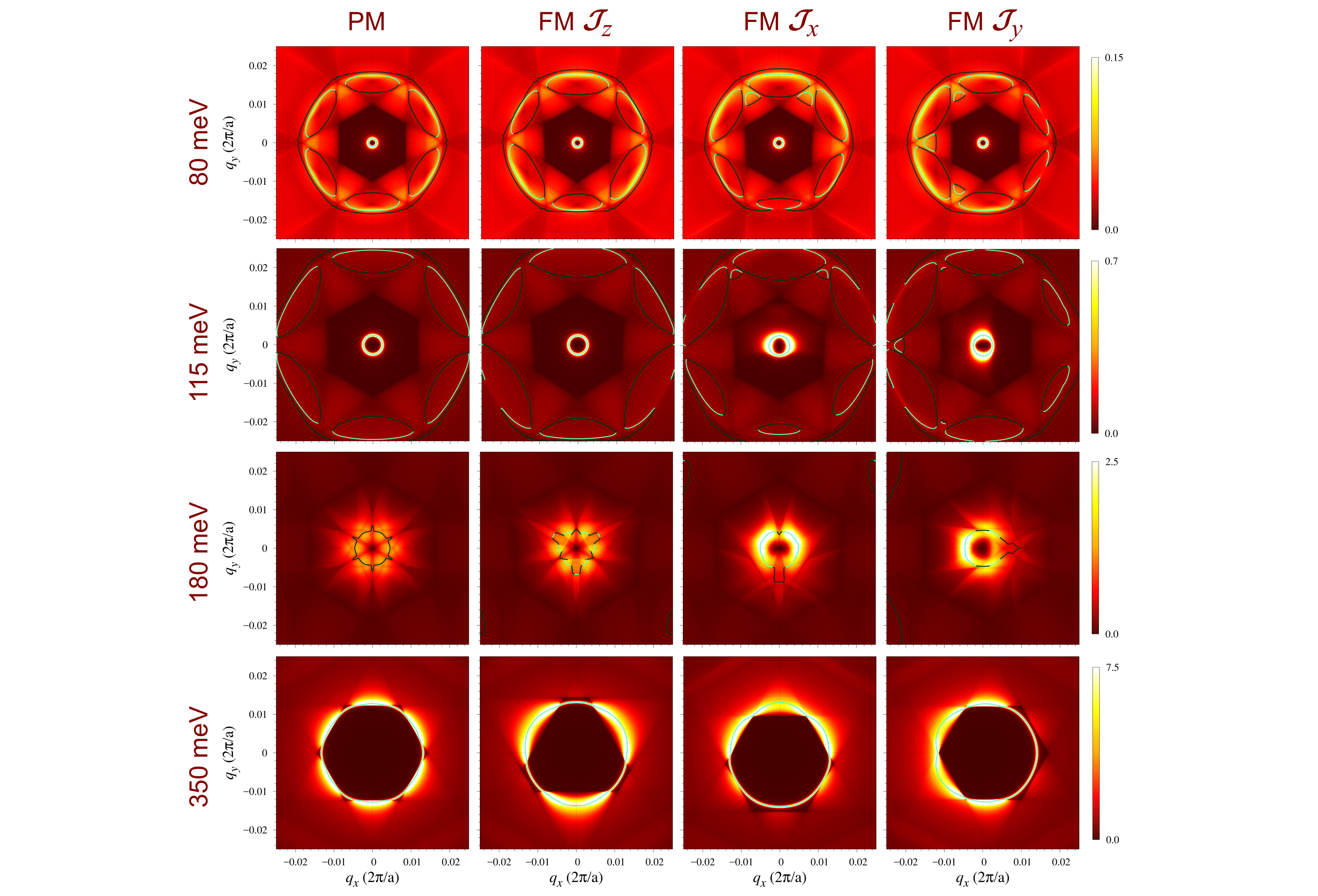}
\caption{Constant-$\omega$ cuts of the energy-loss spectra of the PM and FM BTI with $\mathcal{E}_{\rtm{F}}=\mathcal{E}_{a}$ for the energies indicated in Fig.~\ref{fig8} by white dashed lines. Light and dark green points are the same as in Fig.~\ref{fig8}.}
\label{fig9}
\end{figure*}

In Fig.~\ref{fig8}, we show the energy-loss spectra of BTI in the PM and FM phases for $\mathcal{E}_{\rtm{F}}=\mathcal{E}_{a}$. We start our detailed analysis with the energy-momentum distribution of the loss function of the PM phase. As seen in the upper-left color map of Fig.~\ref{fig8}, the related distribution of $L(\mathbf{q}, \omega)$ has a prominent feature revealing the plasmon mode with a $\sqrt{q}$-like dependence at small \textbf{q}. In this long-wavelength region, the plasmon mode is undamped (the imaginary part of the dielectric function $\epsilon_2$ vanishes), and its dispersion $\omega_p(\mathbf{q})$ is defined by the zeros of the real part of the dielectric function $\epsilon_1$. [Light green points in the figure are a guide for the eye, indicating the roots of $\epsilon_1 (\mathbf{q}, \omega_p(\mathbf{q})) = 0$ that correspond to the positive energy derivative $\xi(\mathbf{q})=\left.\left[\partial \epsilon_1 (\mathbf{q}, \omega)/ \partial \omega\right]^{-1} \right|_{\omega=\omega_p}$, which in the undamped region defines the strength of the plasmon mode.] Here, the plasmon dispersion is practically isotropic, as clearly seen in the $\omega=\rm const$ cuts of the PM loss spectrum for $\omega=80$~meV and 115~meV, see Fig.~\ref{fig9}.

%+++++++++++++++++++++++++++++++++++++++++++++++++++++++++++++++++++++++++++++++++
%+++++++++++++++++++++++++++++++++++++++++++++++++++++++++++++++++++++++++++++++++

At energies slightly above 120~meV the plasmon enters the interbranch electron-hole continuum and acquires a finite spectral width rapidly increasing with $\omega$. Due to the damping, the peak of the loss function shifts from the zeros of $\epsilon_1$ (the light green points). (Note that the stronger the dumping the further the peak may shift.) Nevertheless, the interpretation of the loss spectrum in terms of a well-defined plasmon mode still remains valid for the main peak: the related pole of the inverse dielectric function is near the real axis in the lower half of the complex plane, i.e., close to $\omega_p(\mathbf{q}) -i\gamma_p(\mathbf{q})$, where the finite imaginary part $\gamma_p(\mathbf{q}) = \epsilon_2  (\mathbf{q}, \omega_p(\mathbf{q})) \xi(\mathbf{q})$, which approximately equals the spectral half-width of the plasmon, meets the condition $\gamma_p(\mathbf{q}) \ll \omega_p(\mathbf{q})$.

In the $\omega$ range $\sim130$--200~meV, the plasmon mode encounters a rich interbranch structure. Here, the single-particle excitations with the prominent absorption (seen as bright diverging rays) are strong enough to force $\epsilon_1$ through the Kramers-Kronig relation to cross the real axis with a negative energy derivative (dark green points in Fig.~\ref{fig8}), thereby inducing avoided-crossing gaps in the plasmon dispersion. In Fig.~\ref{fig9}, to illustrate the effect of the single-particle modes we show the behaviour of the loss function over the whole \textbf{q}-plane at an energy of one of the avoided-crossing gaps, see the 180~meV cut of the PM loss spectrum. Then, the plasmon mode extends continuously to $\omega\sim300$~meV and the damping further increases. For $\mathbf{q}\parallel\bar{\Gamma}$-$\bar{M}$ (along $\XU$ in Fig.~\ref{fig8}) it leaves the interbranch continuum and becomes undamped until it re\"{e}nters the continuum at $\omega\sim500$~meV, see the PM loss spectrum in Fig.~\ref{fig8}. For $\mathbf{q}\parallel\bar{\Gamma}$-$\bar{M}$ (along $\YU$),  the plasmon mode does not leave the continuum but for $\omega$ around 350~meV goes very close to the border of the interbranch continuum, with the spectral width being substantially reduced. These features of the plasmon dispersion for $\omega\sim300$--500~meV are clearly seen in the $\omega=\rm const$ cut of the PM loss spectrum at 350~meV, Fig.~\ref{fig9}, as bright thin arcs near the corners of the dark ($\epsilon_2=0$) hexagon (the $\bar{\Gamma}$-$\bar{M}$ lines, $\varphi_{\mathbf{q}}= m\pi/3$ with the integer $m$ varying from 0 to 5) and the fat segments of high intensity around the center of the hexagon sides [the $\bar{\Gamma}$-$\bar{K}$ lines, $\varphi_{\mathbf{q}}= (m+1/2)\pi/3$]. Note also a tangible hexagonal warping of the plasmon-dispersion contour.

Another interesting spectral feature is a linearly dispersing plasmon mode that extends continuously up to $\omega\sim90$~meV for $\mathbf{q}\parallel\bar{\Gamma}$-$\bar{K}$, see the PM case in Fig.~\ref{fig8}. As seen in Fig.~\ref{fig9} (the cut of the PM loss function at $\omega=80$~meV), on the \textbf{q}-plane this damped acoustic mode has a peculiar six-fold symmetric pattern with rather bright arcs disconnected at the $\bar{\Gamma}$-$\bar{M}$ lines, where this plasmon mode disappears. This mode originates from a large difference between the intrabranch electron-hole continua of the inner and outer branches. [Note that in the classical (linear) Rashba model the intrabranch continua are the same for both branches.] In some \textbf{q}-directions, the difference leads to a two-peak structure in the total electron-hole continuum of the intrabranch transitions with a significant dip between the peaks (not shown), which generates an additional $\gamma_p$-positive zero in $\epsilon_1$ within the dip of $\epsilon_2$, resembling the zone-boundary effect~\cite{Foo_PR_1968}. The lower-energy peak of the two-peak structure belongs to the intrabranch continuum of the inner branch, while the higher-energy peak relates to the outer-branch continuum, which is wider in energy and depends stronger on \textbf{q} than the inner-branch continuum. With increasing $\omega$, the shape of the inner- and outer-branch continuum changes, so for $\omega>90$~meV the whole electron-hole intrabranch continuum no longer has the dip, whereby the related $\gamma_p$-positive zeros of $\epsilon_1$ disappear, see the PM loss spectrum for $\mathbf{q}\parallel\bar{\Gamma}$-$\bar{K}$ shown in Fig.~\ref{fig8} \footnote{{At $\omega$ slightly above 100~meV, the peak of the inner-branch continuum is now superimposed on a flattened peak of the outer-branch continuum, whereby the whole intrabranch continuum produces an intense peak in $\epsilon_2$ followed by a shoulder. This shape of $\epsilon_2(\omega)$ gives rise to a $\gamma_p$-positive zero of $\epsilon_1(\omega)$ with a strong damping due to the large $\epsilon_2(\omega)$.}}.

In Fig.~\ref{fig8}, the loss spectrum of the FM BTI magnetized along $\ZU$ looks very similar to the PM spectrum for $\mathbf{q}\parallel\mathbf{\hat x}$ and $\mathbf{q}\parallel\mathbf{\hat y}$. The only notable difference is that now, at relatively high $\omega$, the $\sqrt{q}$-like plasmon mode exhibits a re\"{e}ntrant behavior---leaving the interbranch continuum and becoming undamped---for $\mathbf{q} \parallel \mathbf{\hat y}$ instead of $\mathbf{q}\parallel\mathbf{\hat x}$, as in the PM phase. However, the inspection of the constant-$\omega$ cuts of the FM spectrum (see the FM $\mathcal{J}_z$ column of Fig.~\ref{fig9}) reveals that the $z$-directed field affects tangibly the behavior of the FM loss spectrum over the whole \textbf{q}-plane. As seen in the figure, by modifying the energy-momentum dispersion of the state $\widetilde{\Psi}_{\bm{\rtm{k}}}^{\rtm{BTI}}$ the field reduces the symmetry of the maps so they become three-fold symmetric, like the CECs of the electron spectral function. This also applies to the contours of the plasmon dispersion whose three-fold distortion gets more evident with increasing $\omega$. As follows from Fig.~\ref{fig9}, the changes in the interbranch transitions caused by the field lead to the appearance of the arcs of a highly damped plasmon in the 180~meV cut---the energy in the avoided-crossing gap in the plasmon dispersion of the PM phase, see Fig.~\ref{fig8}. In the 350~meV cut, this results in reshaping the dark hexagon (the $\epsilon_2=0$ area) and, as a consequence, in changing the \textbf{q}-directions along which the plasmon is undamped. Now these directions are at $\varphi_{\mathbf{q}}= 2m\pi/3+\pi/2$ with $m=0$, 1, 2. Note that the acoustic plasmon mode also acquires preferable \textbf{q}-directions, where it is less damped (see the cut of the FM $\mathcal{J}_{z}$ loss spectrum at 80~meV).

 %+++++++++++++++++++++++++++++++++++++++++++++++++++++++++++++++++++++++++++++++++
\begin{figure}[tbp]
\centering
\includegraphics[width=\columnwidth]{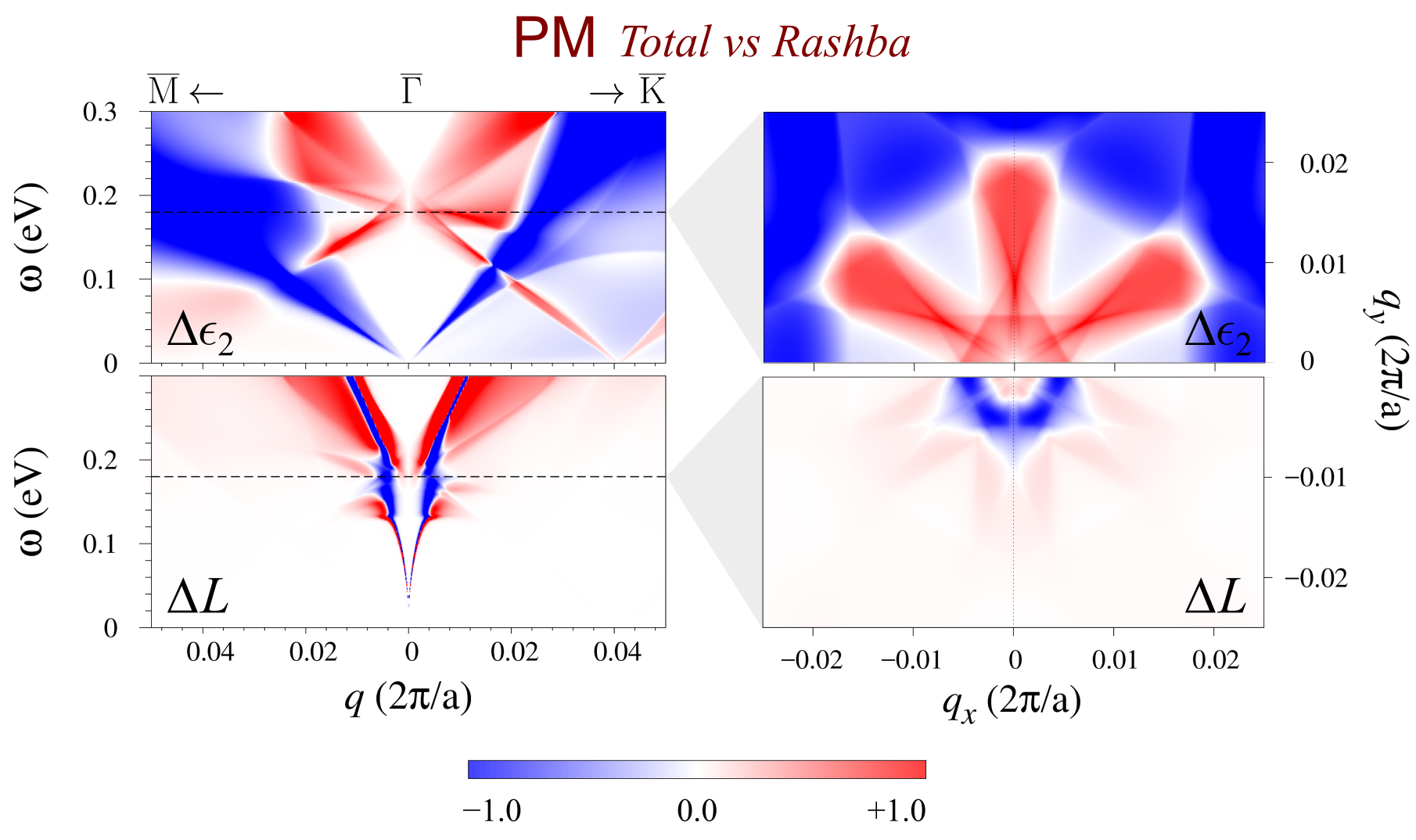}
\caption{The differential spectra $\Delta\epsilon_2 (\omega,\mathbf{q})$ and $\Delta L(\omega,\mathbf{q})$ and  their constant-$\omega=180$~meV cuts for the PM BTI (see text).}
\label{fig10}
\end{figure}
%+++++++++++++++++++++++++++++++++++++++++++++++++++++++++++++++++++++++++++++++++

%+++++++++++++++++++++++++++++++++++++++++++++++++++++++++++++++++++++++++++++++++
\begin{figure*}[tbp]
\centering
\includegraphics[width=\textwidth]{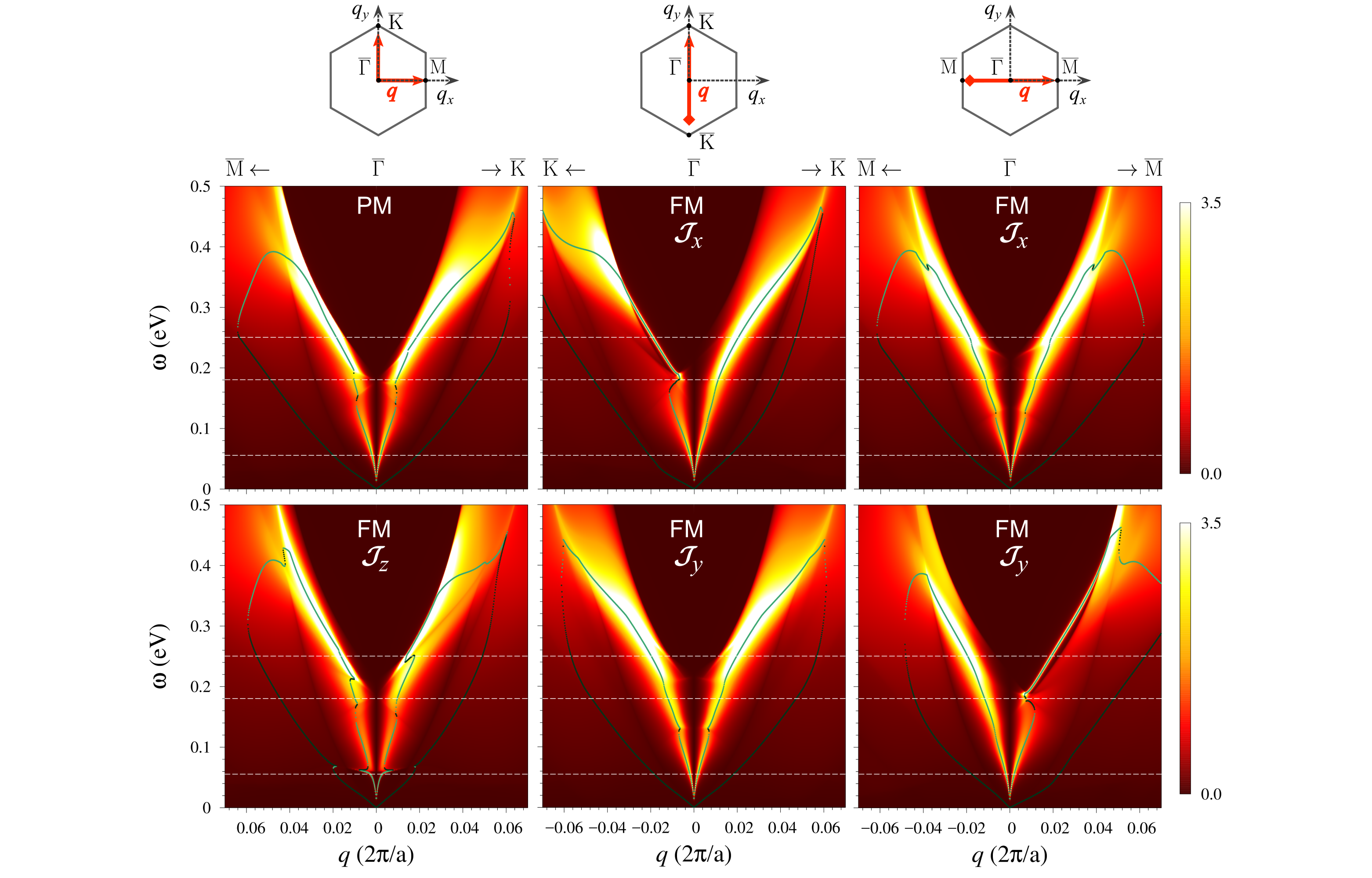}
\caption{Same as in Fig.~\ref{fig8} but for BTI in the PM and FM phases with $\mathcal{E}_{\rtm{F}}=\mathcal{E}_{\rtm{R}}$. The constant $\omega$ cuts of the loss spectra, see Fig.~\ref{fig12}, are for the energies marked by white dashed lines. }
\label{fig11}
\end{figure*}
%+++++++++++++++++++++++++++++++++++++++++++++++++++++++++++++++++++++++++++++++++

The loss spectra of FM BTI with the in-plane magnetization, are strongly asymmetric upon the reversal $\mathbf{q} \to -\mathbf{q}$ for \textbf{q} perpendicular to the field, see the middle and right columns in Fig.~\ref{fig8}. The asymmetry is most clearly manifested in the shift of the minimum of the upper border of the interbranch continuum as well as in the displacement of the meeting points of the bright rays related to the single-particle modes. The field-induced modifications are also seen in the constant-$\omega$ cuts in Fig.~\ref{fig9}: The borders of both the intrabranch continuum (the dark hexagon in the cuts at $\omega=80$, 115, and 180~meV) and the interbranch continuum (the dark hexagon in the 350~meV cut) are shifted by the field in the direction perpendicular to the magnetization.

%+++++++++++++++++++++++++++++++++++++++++++++++++++++++++++++++++++++++++++++++++
\begin{figure*}[tbp]
\centering
\includegraphics[width=\textwidth]{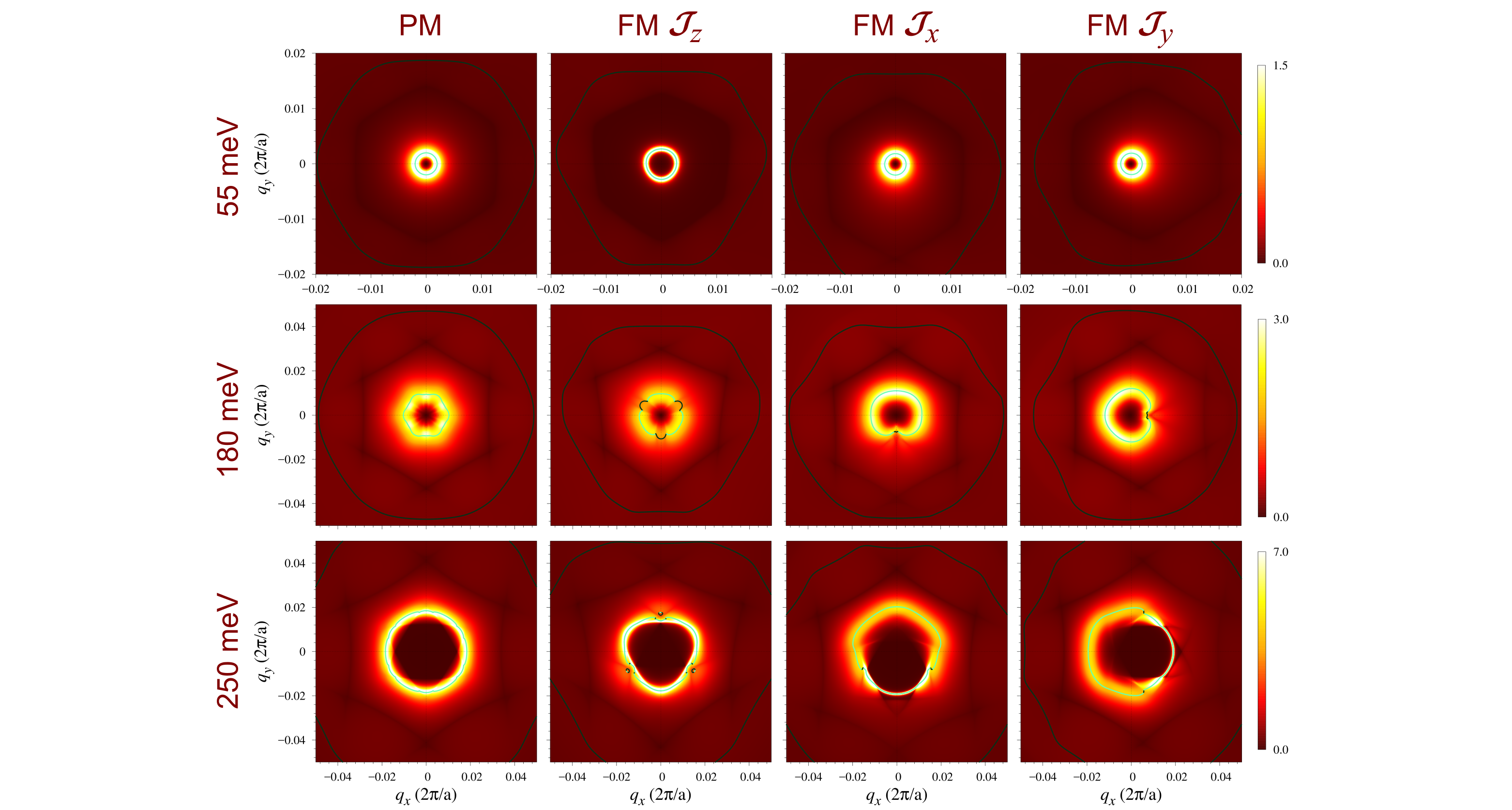}
\caption{Same as in Fig.~\ref{fig9} but for BTI in the PM and FM phases with $\mathcal{E}_{\rtm{F}}=\mathcal{E}_{\rtm{R}}$ and the values of $\omega$ indicated in Fig.~\ref{fig11} by white dashed lines.}
\label{fig12}
\end{figure*}
%+++++++++++++++++++++++++++++++++++++++++++++++++++++++++++++++++++++++++++++++++

The $\sqrt{q}$-like mode at small \textbf{q} perpendicular to the field is rather asymmetric, so it enters the interbranch continuum at different energies for $\mathbf{q}$ and $-\mathbf{q}$, see Fig.~\ref{fig8}. As a consequence, for certain energies slightly below $120$~meV the plasmon remains undamped only for $\varphi_{\mathbf{q}} \sim \varphi_{\bm{\mathcal{J}}} -\pi/2$, where $\varphi_{\bm{\mathcal{J}}}$ is the polar angle of the in-plane field $\bm{\mathcal{J}}$ (e.g., $\varphi_{\bm{\mathcal{J}}}=0$ for $\mathcal{J}_x$ and $\pi/2$ for $\mathcal{J}_y$), while in other \textbf{q}-directions the mode is tangibly damped, see the 115~meV cuts in Fig.~\ref{fig9}. Within the interbranch continuum, the in-plane magnetization affects strongly the plasmon dispersion by modifying the single-particle modes, see the $\omega$ range 120--240~meV of the FM $\mathcal{J}_{x,y}$ loss spectra in Fig.~\ref{fig8}. In the 180~meV cuts in Fig.~\ref{fig9} bright plasmon arcs appear in the avoided-crossing gap of the PM plasmon dispersion. For higher $\omega$, a rather large part of the continuous contour of the plasmon mode now lies in the $\epsilon_2=0$ region, i.e., outside the interbranch continuum because of the shift of the continuum borders, see the cuts of the FM $\mathcal{J}_{x,y}$ loss spectra at $\omega=350$~meV in Fig.~\ref{fig9}. This part of the contour is only slightly warped, while the part that occurs inside the interbranch continuum becomes strongly distorted. Note that the effect of the in-plane field on the acoustic plasmon is that now the peak of the acoustic mode is well defined only in a certain \textbf{q}-sector with $\varphi_{\mathbf{q}}$ varying from $\sim \varphi_{\bm{\mathcal{J}}}$ to $\sim \varphi_{\bm{\mathcal{J}}} + \pi$, while in other \textbf{q}-directions the acoustic mode becomes overdamped.

In order to get an idea of how the locking nonorthogonality manifests itself in the screening properties, for PM BTI we consider the following differential spectra: $\Delta\epsilon_2 (\omega,\mathbf{q}) = \rtm{Im} [\epsilon (\omega,\mathbf{q}) - \epsilon^{\rtm{R}} (\omega,\mathbf{q})]$ and $\Delta L (\omega,\mathbf{q}) = L(\omega,\mathbf{q}) - L^{\rtm{R}} (\omega,\mathbf{q})$ with $L^{\rtm{R}} (\omega,\mathbf{q}) = - \rtm{Im} [1/\epsilon^{\rtm{R}} (\omega,\mathbf{q})]$. Here, the dielectric function $\epsilon^{\rtm{R}}$ is calculated again from Eqs.~(\ref{DF_RPA}) and (\ref{Chi0}) but with the factor $\mathcal{F}^{\lambda\lambda^{\prime}}_{\mathbf{k}, \mathbf{k}+\mathbf{q}} = [1 + \lambda \lambda^{\prime} \cos (\varphi_{\mathbf{k}} - \varphi_{\mathbf{k}+\mathbf{q}}) ] / 2$ taken from the classical (linear) Rashba model. This allows us to highlight the role of the realistic spin structure while keeping the energy-momentum dispersion of $\widetilde{\Psi}_{\bm{\rtm{k}}}^{\rtm{BTI}}$ unchanged.

%+++++++++++++++++++++++++++++++++++++++++++++++++++++++++++++++++++++++++++++++++
\begin{figure*}[tbp]
\centering
\includegraphics[width=\textwidth]{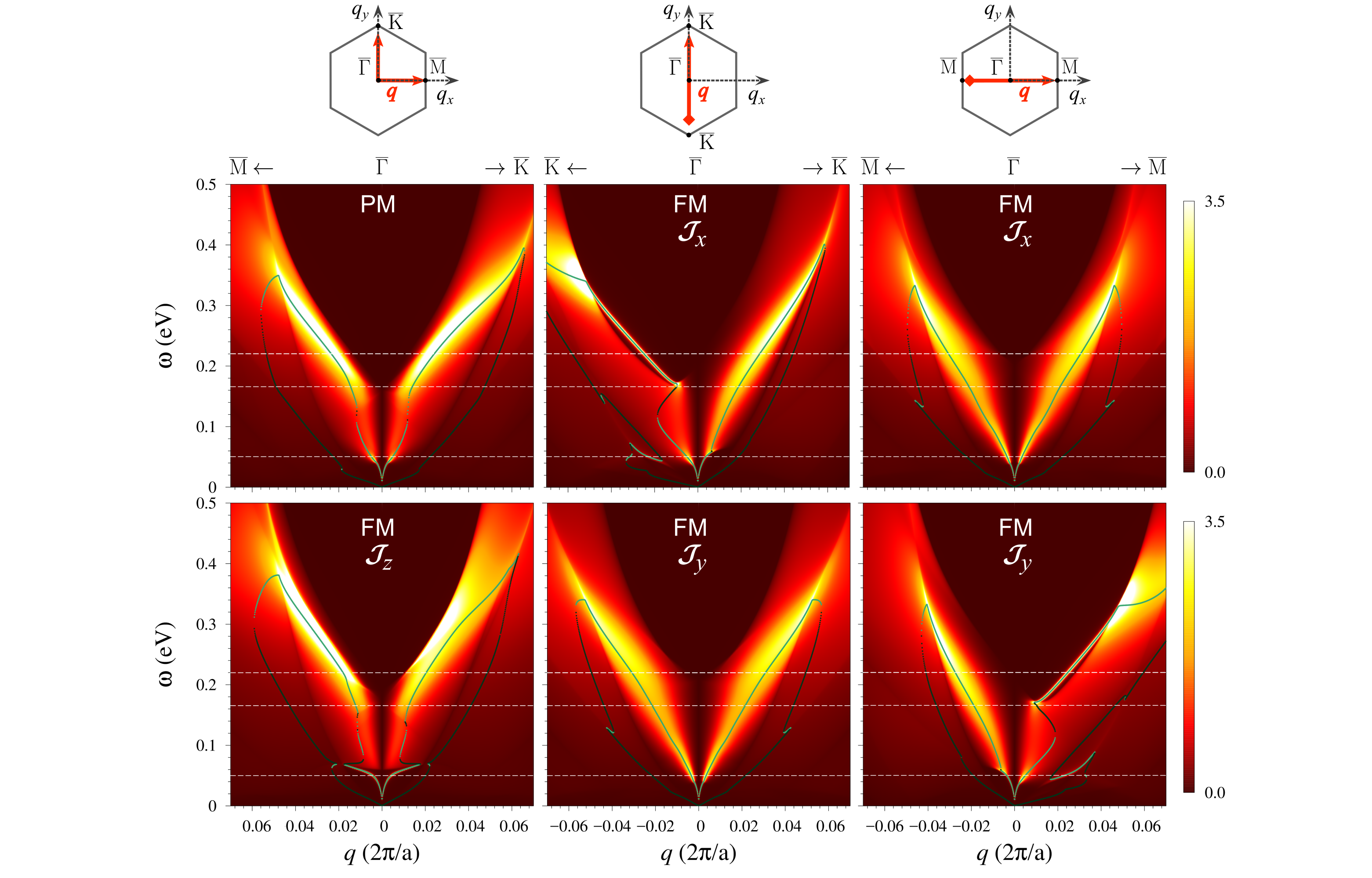}
\caption{Same as in Fig.~\ref{fig11} but for $\mathcal{E}_{\rtm{F}}=\mathcal{E}_{b}$.}
\label{fig13}
\end{figure*}
%+++++++++++++++++++++++++++++++++++++++++++++++++++++++++++++++++++++++++++++++++

In Fig.~\ref{fig10}, we show the differences $\Delta\epsilon_2$ and $\Delta L$ for the $\omega$ range dominated by the transitions between states whose in-plane spin expectation value $\mathbf{S}_{\mathbf{k}} ^{\lambda}$ prevails over the out-of-plane one. As seen in the $\Delta\epsilon_2$ panels of the figure, the absorption is tangibly enhanced by the non-orthogonality in the interbranch continuum (red-shade areas), while in the intrabranch continuum it is noticeably reduced (blue-shade areas). Note that the changes are observed already at small $\omega$. The cut of $\Delta\epsilon_2$ in Fig.~\ref{fig10} shows a non-trivial distribution of the difference over the \textbf{q}-plane at $\omega=180$~meV: within the interbranch continuum the maximum absorption enhancement occurs for $\mathbf{q}\parallel\bar{\Gamma}$-$\bar{K}$.

The effect of the non-orthogonality on the energy-loss function is negligible at small $\omega$ and becomes visible for $\omega$ slightly above 100~meV due to the growing difference between the dispersion of the plasmon modes in $L(\omega,\mathbf{q})$ and $L^{\rtm{R}} (\omega,\mathbf{q})$, see the $\Delta L$ panels in Fig.~\ref{fig10}. The non-orthogonality starts to affect tangibly the loss spectrum when the plasmon modes enter the interbranch continuum. Now, in addition to the position of the main plasmon peaks, the spectral width of the plasmon modes and their interplay with the electron-hole excitations are also affected by the non-orthogonality. The 180~meV cut of  $\Delta L$ in Fig.~\ref{fig10} demonstrates the changes in the plasmon--continuum interplay induced by the non-orthogonal spin-orbit locking: in $L^{\rtm{R}}$ the damped plasmon appears for $\mathbf{q}\parallel \bar{\Gamma}$-$\bar{M}$, while in $L(\omega,\mathbf{q})$ for all \textbf{q} there is an avoided-crossing gap of the plasmon dispersion. Note that at $\omega=180$~meV the electron-hole transitions already involve high-energy states of BTI with the deviation angle $\delta^{\lambda}_{\mathbf{k}}$ exceeding $10^{\circ}$.

Now we analyse the loss spectra of BTI for $\mathcal{E}_{\rtm{F}}=\mathcal{E}_{\rtm{R}}$, which  has only one plasmon mode with a $\sqrt{q}$-like dispersion in the long-wavelength region, Fig.~\ref{fig11}. In the PM phase, the plasmon mode lies entirely in the interbranch continuum and has significantly less avoided-crossing gaps than for $\mathcal{E}_{\rtm{F}}=\mathcal{E}_{a}$.  The intrabranch continuum is formed here by the outer branch only. At small energies, the plasmon dispersion is nearly isotropic, while close to the avoided-crossing gaps it is strongly hexagonally distorted, compare the cuts of the PM loss spectrum at $\omega=55$ and 180~meV in Fig.~\ref{fig12}.

The $z$-directed exchange field induces the gap at $\bar{\Gamma}$ in the band structure of BTI, and, therefore, the plasmon becomes undamped at small \textbf{q}, see the lower-left panel of Fig.~\ref{fig11}. Note also the closeness of the plasmon dispersion to the upper border of the interbranch continuum around 300~meV for $\mathbf{q}\parallel\bar{\Gamma}$-$\bar{K}$. The field-induced symmetry reduction clearly manifests itself already at small energies, see the three-fold patterns shown in the FM $\mathcal{J}_z$ column of Fig.~\ref{fig12}. These patterns also demonstrate that at $\omega=180$~meV the plasmon mode disappears in the \textbf{q}-directions of $\varphi_{\mathbf{q}}= (2m+1/2)\pi/3$ ($m=0$, 1, 2), and that at $\omega=250$~meV the plasmon contour is strongly distorted by the single-particle modes at $\varphi_{\mathbf{q}}= 2m\pi/3 + \pi/2$.

%+++++++++++++++++++++++++++++++++++++++++++++++++++++++++++++++++++++++++++++++++
\begin{figure*}[tbp]
\centering
\includegraphics[width=\textwidth]{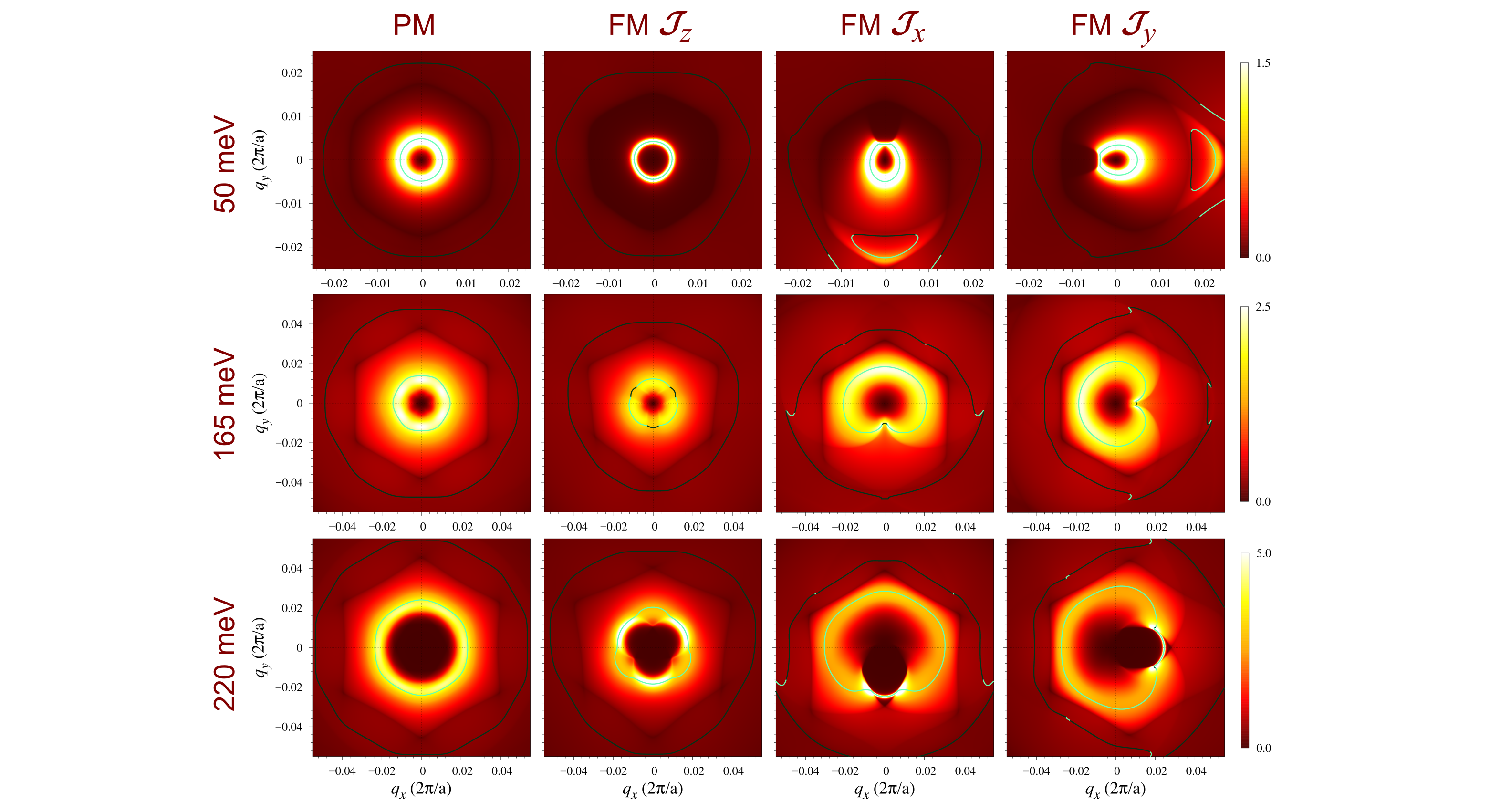}
\caption{Same as in Fig.~\ref{fig12} but for $\mathcal{E}_{\rtm{F}}=\mathcal{E}_{b}$ and the values of $\omega$ indicated in Fig.~\ref{fig13} by white dashed lines.}
\label{fig14}
\end{figure*}
%+++++++++++++++++++++++++++++++++++++++++++++++++++++++++++++++++++++++++++++++++

As seen in Fig.~\ref{fig11}, the in-plane magnetization practically destroys the avoided-crossing gaps: notable gaps are observed only for \textbf{q} close to the direction perpendicular to the exchange field. Similar to the case of $\mathcal{E}_{\rtm{F}}=\mathcal{E}_{a}$, the in-plane magnetization induces a strong asymmetry of the FM $\mathcal{J}_{x,y}$ loss spectra upon the reversal $\mathbf{q} \to -\mathbf{q}$ for \textbf{q} perpendicular to the field. Note also that for these \textbf{q} the plasmon mode leaves the interbranch continuum at $\omega\sim200$~meV and re\"{e}nters it at $\omega\sim350$~meV. Outside the continuum the plasmon is undamped: In the constant-$\omega$ cuts (e.g., at $\omega=250$~meV in Fig.~\ref{fig12}) this manifests itself as bright thin arcs of the plasmon-dispersion contour in the \textbf{q}-sector around $\varphi_{\mathbf{q}}=\varphi_{\bm{\mathcal{J}}} -\pi/2$. As follows from Fig.~\ref{fig12}, the field-induced distortion of the plasmon contour strongly depends on $\omega$. For instance, at $\omega=55$~meV the nearly circular contour slightly shifts as a whole along the \textbf{q}-direction of $\varphi_{\mathbf{q}} = \varphi_{\bm{\mathcal{J}}} -\pi/2$. At $\omega=180$~meV (close to the minimum of the outer border of the intrabranch continuum, see Fig.~\ref{fig11}) the contour strongly bends towards its center near  $\varphi_{\mathbf{q}} = \varphi_{\bm{\mathcal{J}}} -\pi/2$ and acquires a discontinuity at this angle. Finally, at $\omega=250$~meV, the contour is divided into two parts: one of the parts lies mostly outside the interbranch continuum and is only slightly distorted, while the other one, which lies entirely in the continuum, is tangibly warped.

At first glance, the loss spectra for the Fermi level below the Rashba point, $\mathcal{E}_{\rtm{F}}=\mathcal{E}_{b}$, resemble those for $\mathcal{E}_{\rtm{F}}=\mathcal{E}_{\rtm{R}}$, cf. Figs.~\ref{fig11} and~\ref{fig13}. However, there are two crucial differences: (i) the presence of the $\omega$--$\mathbf{q}$ region with $\epsilon_2=0$ at small $\omega$ and \textbf{q} in the loss spectra of both the FM and PM phase and (ii) the appearance of a new plasmon mode upon the application of the in-plane field for \textbf{q} close to the direction perpendicular to the field. This mode is highly damped; it is caused by a gap between the intrabranch and interbranch continua, and, therefore, its dispersion follows closely the upper border of the intrabranch continuum.

As in the case of $\mathcal{E}_{\rtm{F}}=\mathcal{E}_{a}$, in the FM phases the borders of the $\epsilon_2=0$ region at small $\omega$ and \textbf{q} strongly depend on the direction of the field. This allows one to effectively manipulate the damping and the dispersion of the plasmon mode in the long-wavelength region, as illustrated in Figs.~\ref{fig13} and \ref{fig14}. Actually, the $z$-directed field tends to shift to higher energies the lower border of the interbranch continuum at small \textbf{q}, thereby reducing the plasmon linewidth to zero in the interval between the energies at which the plasmon mode enters into the PM or FM continuum. The exchange field along $\ZU$ also tangibly distorts the the plasmon contour making it three-fold symmetric, see the constant-$\omega$ cut of the FM $\mathcal{J}_{z}$ loss spectrum at 55~meV in Fig.~\ref{fig14}.

Regarding the in-plane magnetization, again we observe the field-induced asymmetry between $+\mathbf{q}$ and $-\mathbf{q}$ of the electron-hole continuum for \textbf{q} perpendicular to the magnetization, see the FM $\mathcal{J}_{x,y}$ energy-loss spectra in Fig.~\ref{fig13}. However, here the asymmetry is enhanced by pushing up above the Fermi level a bunch of the electronic states $|\widetilde{\Psi}^{-}_{\mathbf{k}}\rangle$ in a certain \textbf{k}-sector of $\varphi_{\mathbf{k}}$ close to $\varphi_{\bm{\mathcal{J}}} - \pi/2$, see Figs.~\ref{fig2} and \ref{fig4}.  In the long-wavelength region, at $\omega$ close to the entrance of the plasmon mode into the interbranch continuum this leads to a strong distortion of the plasmon contour and to the anisotropic plasmon linewidth which, unlike the $\mathcal{E}_{\rtm{F}}=\mathcal{E}_{a}$ case, vanishes for $\varphi_{\mathbf{q}} \sim \varphi_{\bm{\mathcal{J}}} +\pi/2$, compare the 50~meV cuts in Fig.~\ref{fig14} and the 80~meV cuts in Fig.~\ref{fig9}.  Also, this gives rise to an additional plasmon mode seen as a rather thick bright arc in the \textbf{q}-sector around $\varphi_{\bm{\mathcal{J}}} - \pi/2$ in the 50~meV cut in Fig.~\ref{fig14}. Note that the higher-energy cuts of the FM $\mathcal{J}_{x,y}$ energy-loss spectra for $\mathcal{E}_{\rtm{F}}=\mathcal{E}_{b}$ (at $\omega=165$ and 220~meV in Fig.~\ref{fig14})  resemble those for $\mathcal{E}_{\rtm{F}}=\mathcal{E}_{\rtm{R}}$ (at $\omega=180$ and 250~meV in Fig.~\ref{fig12}).

In short, the above analysis of the BTI loss function shows that in the PM phase the hexagonal warping of the $\widetilde{\Psi}_{\bm{\rtm{k}}}^{\rtm{BTI}}$ CECs (mainly of the outer contours) produces the six-fold symmetric electron-hole continuum with a rich structure. The resulting plasmon--continuum interplay leads to the avoided-crossing gaps in the plasmon dispersion. The plasmon contours also are hexagonally distorted, especially when they overlap with the interbranch continuum. The significant difference in the structure of the intrabranch continua of the inner and outer branches gives rise to the acoustic plasmon mode in some \textbf{q}-directions. The comparison with the orthogonal coupling of the classical Rashba model shows that even a rather small non-orthogonality tangibly enhances the interbranch absorption and reduces the intrabranch absorption, which in turn alters the behavior of the plasmon mode. The imposition of the exchange field affects all the mentioned features of the PM loss spectrum by modifying the energy-momentum dispersion and spin structure of $\widetilde{\Psi}_ {\bm{\rtm{k}}} ^{\rtm{BTI}}$. The field reduces the symmetry of the spectrum, changes the plasmon--continuum interplay and the plasmon dispersion, and for certain $\omega$ has a strong impact on the damping of the plasmon mode, mainly favoring the mode propagation along the \textbf{q}-direction perpendicular to the field.

%+++++++++++++++++++++++++++++++++++++++++++++++++++++++++++++++++++++++++++++++++
\begin{figure*}[tbp]
\centering
\includegraphics[width=\textwidth]{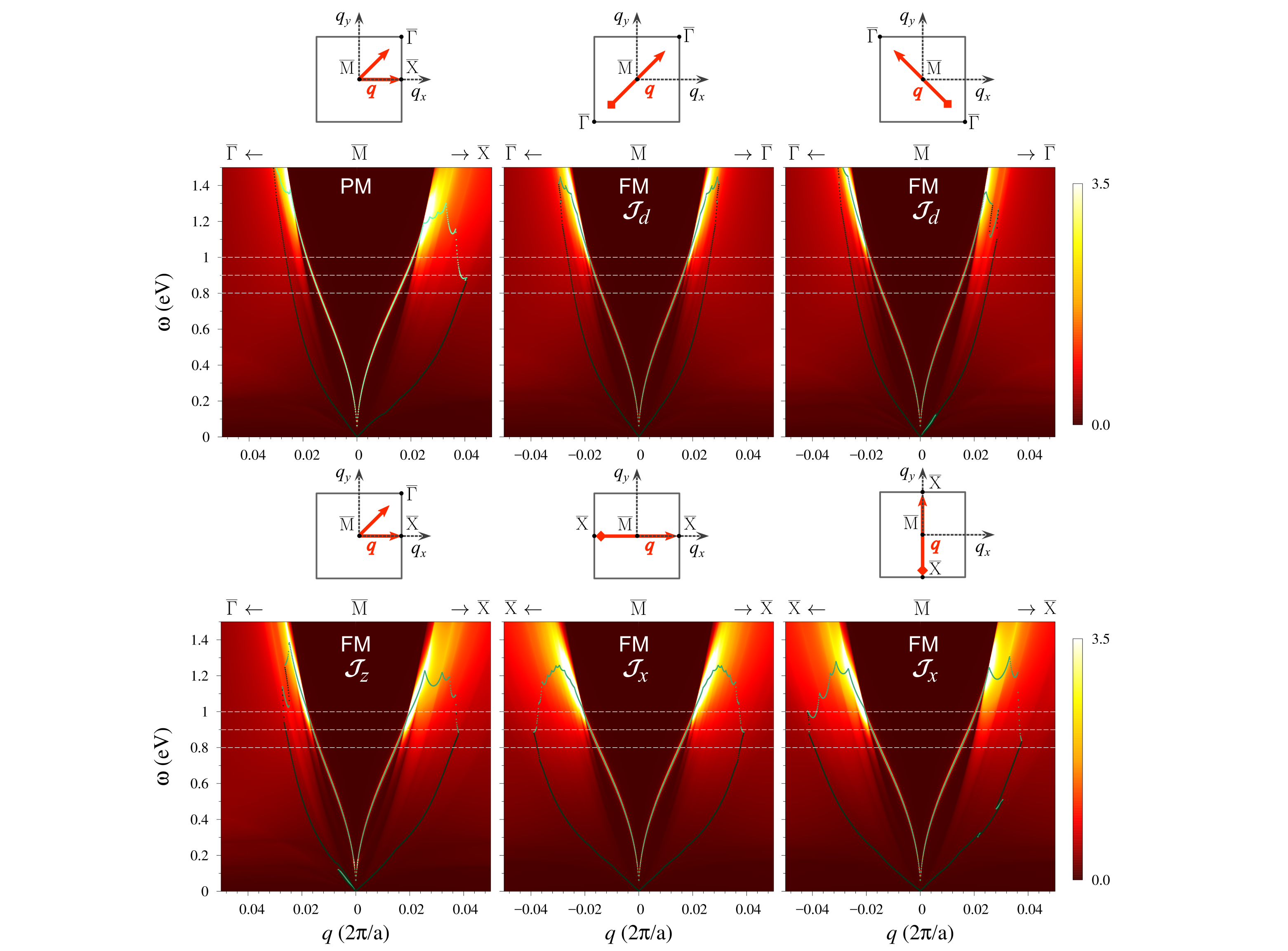}
\caption{Energy-loss spectra of PM and FM TRS with $\mathcal{E}_{\rtm{F}}=\mathcal{E}_{a}$ along the $\mathbf{q}$-lines indicated at the top of each color map. White dashed lines show the energies at which the constant-$\omega$ cuts of the loss spectra are calculated, see Fig.~\ref{fig16}. Light and dark green points are the same as in the figures of Sec.~\ref{C3v_plasmons}: the zeros of the real part of the dielectric function with positive and negative energy derivative of $\epsilon_1$, respectively.}
\label{fig15}
\end{figure*}
%+++++++++++++++++++++++++++++++++++++++++++++++++++++++++++++++++++++++++++++++++

\subsection{\label{C4v_plasmons}Cubic symmetry: TbRh$_2$Si$_2$}

The TRS model differs from BTI by a considerably weaker SOI and much stronger warping of the scalar-relativistic CECs of the $\widetilde{\Psi}_{\bm{\rtm{k}}} ^{\rtm{TRS}}$ contour. Figure~\ref{fig15} shows the energy-loss spectra of TRS in the PM and FM phases for $\mathcal{E}_{\rtm{F}}=\mathcal{E}_{a}$. [Only one position of the Fermi level is considered for TRS, see Fig.~\ref{fig5}.] Already the PM loss spectrum vividly demonstrates a crucial difference from BTI: there is only one plasmon mode, which is undamped over the wide $\omega$ range: the plasmon leaves the interbranch continuum already at $\omega\sim100$~meV and re\"{e}nters it only above 1100~meV, Fig.~\ref{fig15}.

%+++++++++++++++++++++++++++++++++++++++++++++++++++++++++++++++++++++++++++++++++
\begin{figure*}[tbp]
\centering
\includegraphics[width=\textwidth]{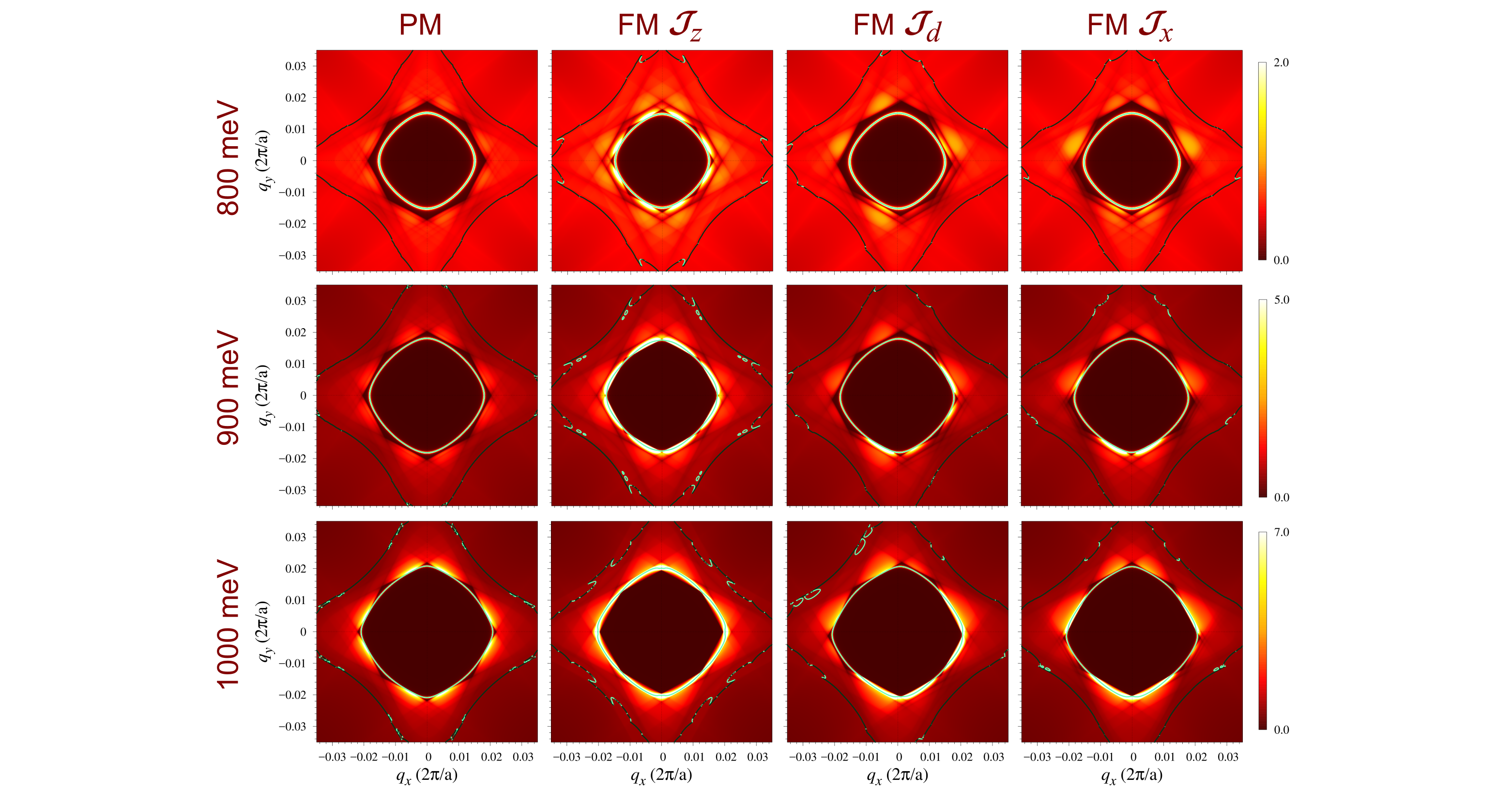}
\caption{Constant-$\omega$ cuts of the energy-loss spectra of the magnetic and non-magnetic TRS model system with $\mathcal{E}_{\rtm{F}}=\mathcal{E}_{a}$ for the energies indicated in Fig.~\ref{fig15} by white dashed lines. Light and dark green points are the same as in Fig.~\ref{fig15}.}
\label{fig16}
\end{figure*}
%+++++++++++++++++++++++++++++++++++++++++++++++++++++++++++++++++++++++++++++++++

The overall pattern of the PM spectrum resembles a classical 2D jellium screening because owing to the weak spin-orbit coupling the interbranch continuum only slightly extends the upper border of the intrabranch continuum toward higher $\omega$. However, the quadrangular star shape of the Fermi contour and the unique spin-momentum locking in $\widetilde{\Psi}_{\bm{\rtm{k}}} ^{\rtm{TRS}}$ (see Fig.~\ref{fig6}) give rise to a non-trivial four-fold geometry and fine structure of the constant-$\omega$ cuts of the PM loss function, see, e.g., polygonal fringes surrounding the dark octagon (the $\epsilon_2=0$ region) in the PM column of Fig.~\ref{fig16}. The rounded square shape of the plasmon contour is clearly seen for all the cuts. The contour is slightly distorted at $\omega=1000$~meV because it approaches the border of the octagon.

\begin{figure}[tbp]
\centering
\includegraphics[width=\columnwidth]{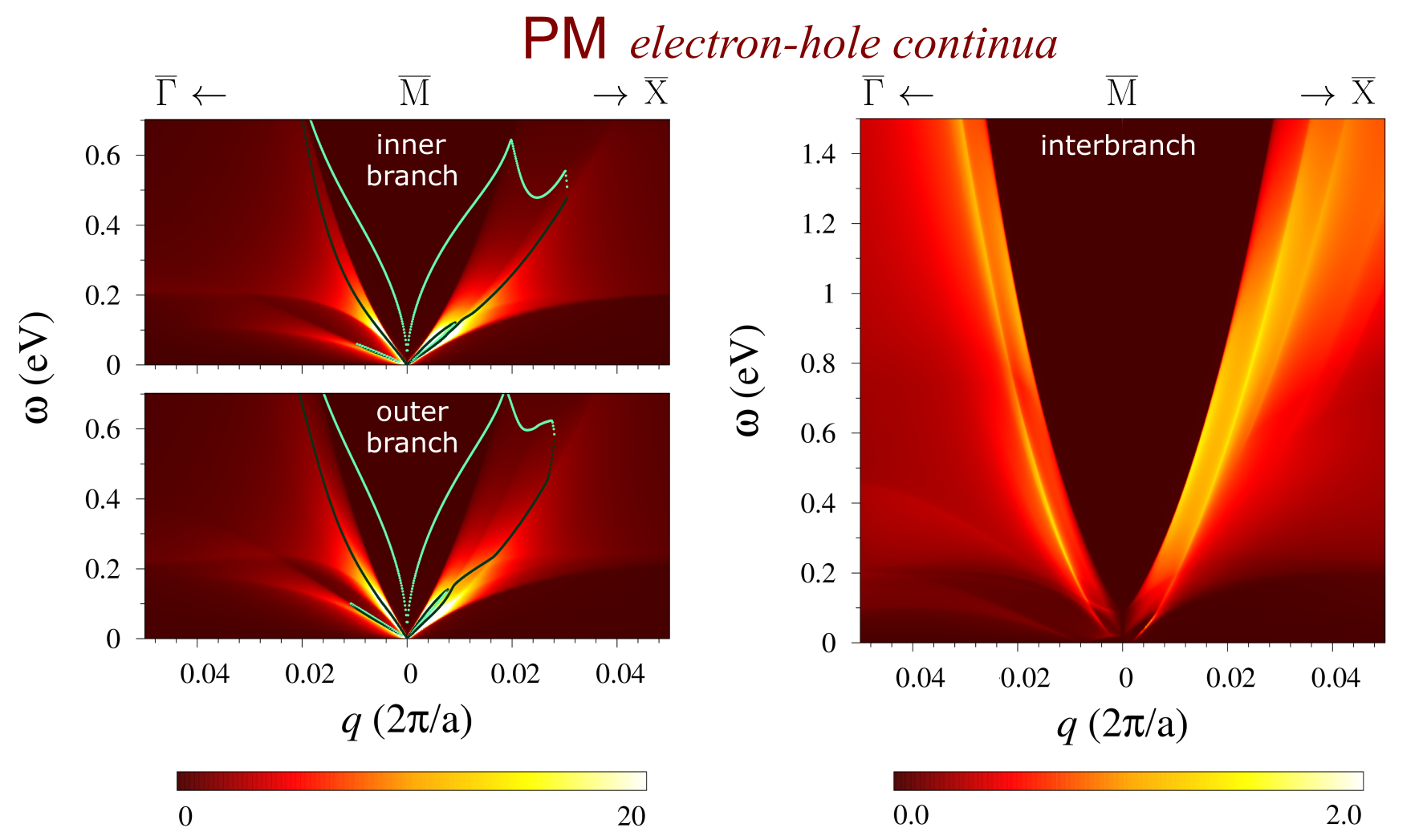}
\caption{The spectra of the intrabranch continua of the inner and outer branch (left panels) as well as the interbranch continuum (right panel) for PM TRS. Green points are zeros of $\epsilon_1$ that is linked through the Kramers-Kronig relation to the partial $\epsilon_2$ formed by only the inner- or outer-branch continuum. As before, the shade of green encodes the sign of the energy derivative of $\epsilon_1$ at a given zero.}
\label{fig17}
\end{figure}

At low $\omega$, the strong four-fold warping of the CECs of $\widetilde{\Psi}_{\bm{\rtm{k}}} ^{\rtm{TRS}}$  manifests itself in a peculiar shape of both intrabranch continua, see the left panels of Fig.~\ref{fig17}, where the light green points show that because $\epsilon_2$ is restricted to either inner or outer intrabranch continuum there appears only a linearly dispersing collective mode in addition to the main plasmon mode similar to Ref.~\cite{LeBlanc_PRB_2014}. However, a deviation between the intrabranch continua caused by the small spin-orbit splitting in TRS turns out to be sufficient to yield the total $\epsilon_2$ such that an additional $\gamma_p$-positive zero in $\epsilon_1$ does not arise.

Now we turn to FM TRS magnetized along $\ZU$. The exchange field strongly enhances the splitting of $\widetilde{\Psi}_{\bm{\rtm{k}}} ^{\rtm{TRS}}$, Fig.~\ref{fig5}, thereby shifting to higher energies the upper border of the interbranch continuum. This reduces the $\omega$ range of the undamped plasmon by $\sim300$~meV, see the FM $\mathcal{J}_{z}$ loss spectrum in Fig.~\ref{fig15}. Actually, now the plasmon mode leaves the continuum at $\omega\sim180$~meV and, as follows from the cuts shown in the FM $\mathcal{J}_{z}$ column of Fig.~\ref{fig16}, already at $\omega=800$~meV the plasmon contour preserving its rounded square shape touches the border of the continuum. Then, at 900~meV the plasmon remains undamped only at $\varphi_{\mathbf{q}} = m\pi/8$ with the integer $m$ varying from 0 to 8, and at 1000~meV it is entirely in the continuum. Note that under the $z$-directed field the cuts preserve their four-fold symmetry.

The $z$-directed exchange field significantly decreases the magnitude of the vector $\mathbf{S}^{\spr \lambda}_{\mathbf{k}}$ (see Fig.~\ref{fig6}) and gives rise to a sizable out-of-plane spin component of the opposite sign for the inner- and outer branch of $\widetilde{\Psi}_ {\bm{\rtm{k}}} ^{\rtm{TRS}}$, see Ref.~\cite{Usachov_PRL_2020}. This substantially reduces the absorption within the interbranch continuum. The field along $\ZU$ also enhances the difference between the inner- and outer-branch continua, so now for $\mathbf{q} \parallel \bar{M}$-$\bar{\Gamma}$ the resulting shape of $\epsilon_2$ brings about a linearly dispersing damped plasmon mode at small $\omega$, see the FM $\mathcal{J}_{z}$ spectrum in Fig.~\ref{fig15}.

%+++++++++++++++++++++++++++++++++++++++++++++++++++++++++++++++++++++++++++++++++
%+++++++++++++++++++++++++++++++++++++++++++++++++++++++++++++++++++++++++++++++++

 %+++++++++++++++++++++++++++++++++++++++++++++++++++++++++++++++++++++++++++++++++
\begin{figure}[tbp]
\centering
\includegraphics[width=\columnwidth]{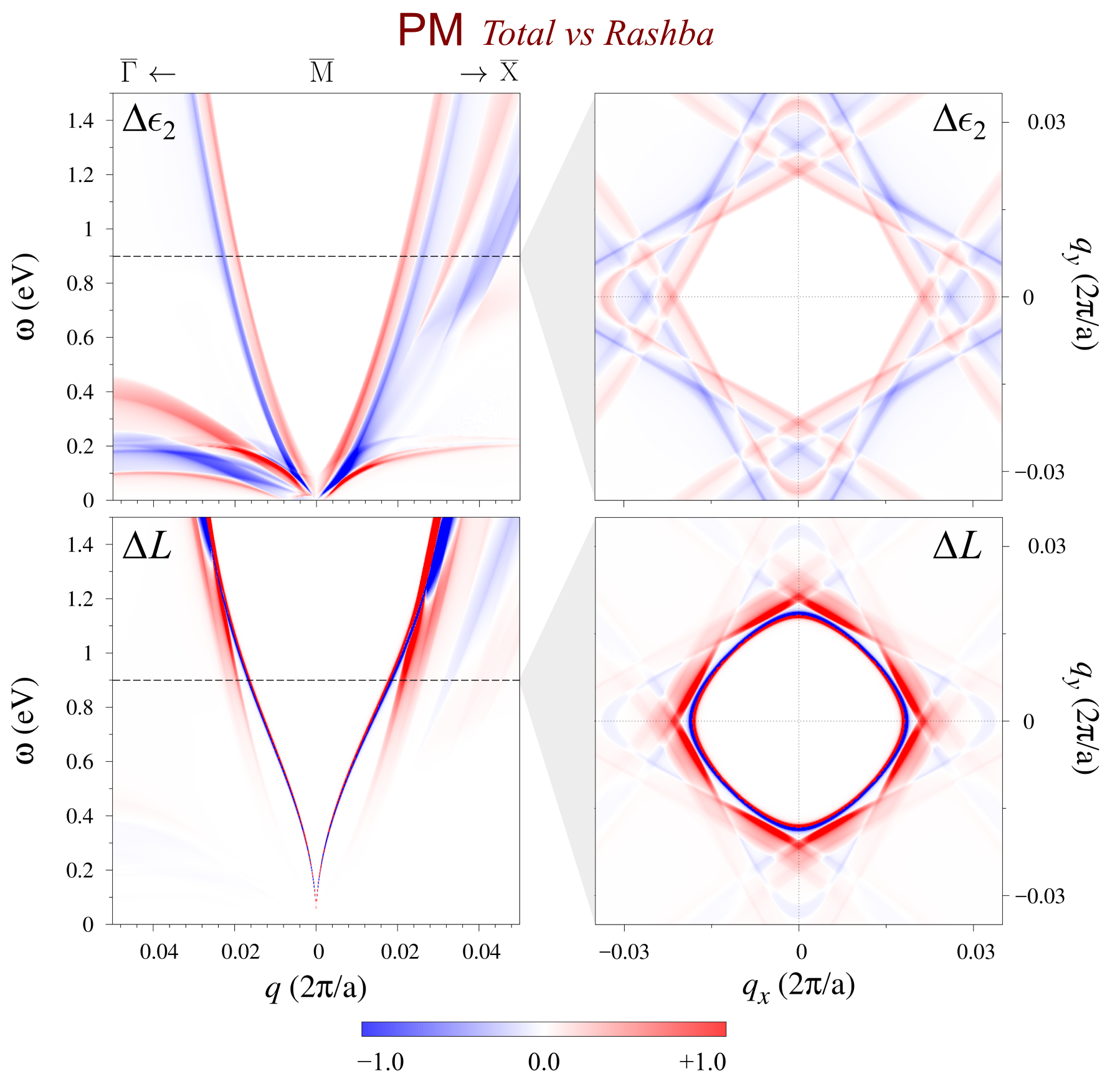}
\caption{The differential spectra $\Delta\epsilon_2 (\omega,\mathbf{q})$ and $\Delta L(\omega,\mathbf{q})$ and their constant-$\omega=900$~meV cuts for PM TRS.}
\label{fig18}
\end{figure}
%+++++++++++++++++++++++++++++++++++++++++++++++++++++++++++++++++++++++++++++++++

As in BTI, the in-plane magnetization induces an asymmetry of the energy-loss spectra upon the reversal $\mathbf{q} \to -\mathbf{q}$ for \textbf{q} perpendicular to the field, though not as prominent as in BTI. The FM $\mathcal{J}_{d,x}$ maps of Fig.~\ref{fig15} look similar to the PM color map, except rather fine details as, e.g., the entrance of the plasmon mode in the interbranch continuum at different $\omega$ above 1000~meV for $+\mathbf{q}$ and $-\mathbf{q}$ and the appearance of a linearly dispersing overdamped mode at low $\omega$ in the FM $\mathcal{J}_{d}$ spectrum for \textbf{q} perpendicular to the field. In that sense, the constant-$\omega$ cuts in Fig.~\ref{fig16} most clearly illustrate the effect of the in-plane field on the loss function of TRS. In particular, unlike the shift of the slightly deformed dark hexagon (the $\epsilon_2=0$ region) observed in the FM BTI loss-function cuts (at 80, 115 or 350~meV in Fig.~\ref{fig9}), the dark octagon is strongly deformed. Outside the deformed octagon, the polygonal fringe pattern noticeably changes, and inside the octagon the plasmon contour is tangibly distorted. The cuts at 900 and 1000~meV  in Fig.~\ref{fig16} demonstrate that the contour further distorts when some of its parts overlap with the electron-hole continuum. Thereby this mode is damped in the \textbf{q}-sectors where it is undamped in the PM phase.

The  effect of the spin structure on the loss function of PM TRS stems from $\widetilde{\Psi}_ {\bm{\rtm{k}}} ^{\rtm{TRS}}$ having only the in-plane spin component $\mathbf{S}^{\spr \lambda}_{\mathbf{k}}$. This means that for the whole $\omega$ range the differential spectra $\Delta\epsilon_2$ and $\Delta L$ in Fig.~\ref{fig18} are exclusively due to the non-Rashba behaviour of $\mathbf{S}^{\spr \lambda}_{\mathbf{k}}$, specifically due to the triple winding of the in-plane spin. Comparing the $\omega$-\textbf{q} map of $\Delta\epsilon_2$ with Fig.~\ref{fig17}, we infer that, similar to the non-orthogonality in BTI, the triple winding tends to increase the probability of  the interbranch transitions (red-shade areas) and to decrease the probability of the intrabranch transitions (blue-shade areas). For the most part, the intrabranch continuum is affected by the triple winding near the upper border of the continuum and in the region of the  warping-induced features discussed above in connection with the appearance of the linearly dispersing modes (see Fig.~\ref{fig17}). The  effect of the non-Rashba locking on $\epsilon_2$ is illustrated in Fig.~\ref{fig18} by the $\omega=900$~meV cut of $\Delta\epsilon_2$, which shows the \textbf{q}-distribution of the modifications due to the triple winding as the polygonal fringe fine structure surrounding the octagon, highlighting the role of the inter- and intrabranch transitions.

The $\omega$-\textbf{q} map of $\Delta L$ in Fig.~\ref{fig18} demonstrates that the loss function is tangibly affected by the change from the Rashba orthogonal locking to the triple winding even if the plasmon occurs rather far from the border of the interbranch continuum. As in BTI, the strongest changes happen when the plasmon enters the continuum. The dispersion itself also undergoes changes; at $\omega\gtrsim500$~meV the slope of the plasmon-dispersion curve in $L(\omega, \mathbf{q})$ becomes visibly larger than in $L^{\rtm{R}}(\omega, \mathbf{q})$. In the $\omega=900$~meV cut of $\Delta L$ this manifests itself as a smaller rounded square---the red contour of the plasmon inside the octagon in Fig.~\ref{fig18}. Note that this octagon is bordered by the bright red stripes reflecting the enhanced contribution of the interbranch transitions to the loss function near the border of the whole electron-hole continuum.

For TRS, we thus have found that the small spin-orbit splitting and the strong warping of both branches of $\widetilde{\Psi}_ {\bm{\rtm{k}}} ^{\rtm{TRS}}$ lead to similar intrabranch-continuum structures of the inner and outer branches, each of which giving rise to a linearly dispersing plasmon mode in addition to the main mode. In real life, when both branches are present, this feature appears in the TRS loss function when the exchange field is introduced. Compared with the Rashba single winding of the in-plane spin, the triple winding generates a finer fringe pattern coming from  the electron-hole continuum and is responsible for an increase in the intrabranch and a decrease in the interbranch absorption. In other words, it acts similar to the few-degree deviation of the locking angle from orthogonality inherent in BTI.

\section{Conclusions}

To summarize, we have developed a minimal two-band model for magnetic 2D electronic systems with spin-orbit coupling based on a fully \ai\ relativistic \kp\ perturbation framework. The \kp\ expansion includes high orders of the crystal momentum $\mathbf{k}$ and accurately reproduces non-Rashba spin-momentum locking patterns coming from \ai\ calculations. This allows a straightforward quantitative treatment of the coupling between the exchange magnetic field and the real spin degrees of freedom in spin–orbit split 2D states.

We applied this model to two prototypical systems of different symmetry---the lowest conduction-band state of a hexagonal BiTeI trilayer and the Rh-derived surface state at the Si-terminated (001) surface of cubic TbRh$_2$Si$2$. The use of the higher-order \kp\ expansion enabled us to account for the warping of the constant-energy contours and the intricate spin textures in these systems, which is essential for an accurate description of the collective modes in the dielectric response.

We analyzed the modification of the electron energy and spin structure induced by the exchange magnetic field of various orientation and demonstrated its impact on the plasmon dispersion and damping. For BTI, the out-of-plane exchange field breaks the sixfold rotational symmetry of the paramagnetic phase and imposes a threefold symmetric distortion of the Fermi contours, accompanied by a change in the plasmon dispersion. In contrast, in TRS the fourfold symmetry of the contours is preserved under the same field orientation, while the in-plane spin retains its triple winding with a reduced magnitude and the appearance of an opposite-sign out-of-plane spin component.

When the in-plane exchange field is applied, both materials exhibit a pronounced asymmetry of the energy spectra upon the reversal $\mathbf{k}\rightarrow-\mathbf{k}$ for $\mathbf{k}$ perpendicular to the field. In BTI this manifests as a shift of the Rashba point away from $\mathbf{k}=0$ and opposite displacements of the inner and outer Fermi contours, resulting in a strong anisotropy of the plasmon mode and a direction-dependent damping. In TRS, the in-plane field drastically modifies the triple-winding spin structure, eventually aligning the in-plane spin with the magnetization direction and substantially changing the matrix elements.

The plasmonic response of both systems is found to be highly sensitive to the spin texture, and understanding its field-induced transformation requires a defiled knowledge of the wave functions. The non-orthogonal spin-momentum locking in BTI enhances the probability of interbranch transitions and suppressed the intrabranch ones, whereas in TRS the triple winding has the opposite effect. This contrasting behavior modifies the structure of the electron–hole continuum and, consequently, the plasmon linewidth and dispersion. The exchange field adds more aspects to this interplay, providing an efficient means to tune the plasmon propagation direction, anisotropy, and damping by changing the orientation and magnitude of the magnetization.

In summary, our study demonstrates that a high-order relativistic \kp\ treatment of spin–orbit coupled 2D systems is essential for a quantitative description of their screening properties in the presence of magnetic exchange interaction. The revealed mechanisms of plasmon manipulation via the interplay between SOI and magnetization open perspectives for designing magnetically tunable plasmonic devices and for exploring collective excitations in spin-orbit materials beyond the conventional Rashba paradigm.

\begin{acknowledgments}
This work was supported by the Department of Education of the Basque Government, grant No.~IT1164-19, and by the Spanish MCIU/AEI/10.13039/501100011033, projects No.~PID2022-139230NB-I00 and PID2022-138750NB-C22.
\end{acknowledgments}

\end{document}